\documentclass[iop]{emulateapj}
\def\gf{$gf$}
\def\teff{$T_{\rm eff}$}
\def\logg{log $g$}
\def\c12{$^{12}$C/$^{13}$C}

\begin{document}

\title{Chemical Abundances for Evolved stars in M5: Lithium through Thorium\altaffilmark{8}}

\author{David K. Lai{\altaffilmark{1,2}}, Graeme
  H. Smith{\altaffilmark{1}}, Michael
  Bolte{\altaffilmark{1}}, 
  Jennifer A. Johnson{\altaffilmark{3}}, Sara
  Lucatello{\altaffilmark{4,5,6}}, Robert P. Kraft{\altaffilmark{1}} and 
  Christopher Sneden{\altaffilmark{7}} }
\altaffiltext{1}{UCO Lick/Department of Astronomy and Astrophysics, University
  of California, Santa Cruz, CA 95064; david@ucolick.org,
  bolte@ucolick.org, graeme@ucolick.org, kraft@ucolick.org}
\altaffiltext{2}{NSF Astronomy and Astrophysics Postdoctoral Fellow}
\altaffiltext{3}{Department of Astronomy, Ohio State University, 140
  W. 18th Ave., Columbus, OH 43210; jaj@astronomy.ohio-state.edu.}
\altaffiltext{4}{Osservatorio Astronomico di Padova, Vicolo dell'Osservatorio 5, 35122 Padua, Italy; sara.lucatello@oapd.inaf.it.}
\altaffiltext{5}{Excellence Cluster Universe, Technische Universit\"at M\"unchen, 
Boltzmannstr. 2, D-85748, Garching, Germany}
\altaffiltext{6}{Max-Planck-Institut f\"ur Astrophysik, Karl-Schwarzschild-Str. 1
85741 Garching, Germany}
\altaffiltext{7}{Department of Astronomy and McDonald Observatory, The
  University of Texas, Austin, TX 78712, chris@verdi.as.utexas.edu}
\altaffiltext{8}{The data presented herein were obtained at the 
W.~M Keck Observatory, which is operated as a scientific 
partnership among the California Institute of Technology, the University of California and the National Aeronautics 
and Space Administration. The Observatory was made possible by the generous financial support of the W.~M. Keck Foundation.} 

\begin{abstract}

We present analysis of high-resolution spectra of a sample of stars in
the globular cluster M5 (NGC 5904). The sample includes stars from the
red giant branch (seven stars), the red horizontal branch (two stars),
and the asymptotic giant branch (eight stars), with effective
temperatures ranging from 4000 K to 6100 K. Spectra were obtained with
the HIRES spectrometer on the Keck I telescope, with a wavelength
coverage from 3700 \AA{} to 7950 \AA{} for the HB and AGB sample, and
5300 \AA{} to 7600 \AA{} for the majority of the RGB sample. We 
find offsets of some abundance ratios between the AGB and the RGB
branches. However, these discrepancies appear to be due to analysis effects,
and indicate that caution must be exerted when directly comparing abundance ratios between
different evolutionary branches. We find the expected signatures of
pollution from material enriched in the products of the hot hydrogen
burning cycles such as the CNO, Ne$-$Na, and Mg$-$Al cycles, but no
significant differences within these signatures among the three stellar
evolutionary branches especially when considering the analysis
offsets. We are also able to measure an assortment of neutron-capture
element abundances, from Sr to Th, in the cluster. We find that the
neutron-capture signature for all stars is the same, and shows a
predominately r-process origin. However, we also see
evidence of a small but consistent extra $s$-process signature that is
not tied to the light-element variations, pointing to a pre-enrichment
of this material in the protocluster gas.

\end{abstract}


\section{Introduction}

Among globular clusters of the northern sky, M5 is one of the nearest,
and the element abundance patterns among its member stars have
received considerable attention. On the basis of its observed
proper motion, M5 actually appears to be an outer halo globular
cluster on an eccentric orbit with a large apogalactic distance of
$\sim 60$ kpc \citep{scholz96}. It is one of
the most metal-rich globular clusters of the outer Galactic halo, with
${\rm [Fe/H]} = -1.34 \pm 0.09$ \citep{carretta09c}.

M5 was one of the first globular clusters in which a
sub-population of red giant branch (RGB) stars whose spectra exhibit
enhanced $\lambda$4215 CN bands were discovered via DDO photometry
\citep{osborn71, hesser77, pike78}. The CN anomalies in M5 have
been traced lower down the giant branch \citep{briley92} and to the
base of the RGB \citep{cbs02}. Abundance variations of O, Na, and Al
also exist among the RGB stars (e.g. \citealt{norris83,ivans01,
  yong08, yong08b, carretta09a, carretta09b}), and inhomogeneities in
Na abundance have been traced from the tip of the RGB to the main
sequence turnoff by \citet{rc03}.  In all these respects, the
abundance inhomogeneities found among RGB stars in M5 appear to be
typical of the broad patterns found in other globular clusters of the
Milky Way (e.g, \citealt{carretta04, gratton01, bragaglia10}) and also
the Local Group \citep{mucciarelli09}.

However, abundance anomalies among the asymptotic giant branch (AGB)
stars of globular clusters, including M5, have not been as well
studied as those on the RGB. In color-magnitude diagrams (CMDs) of M5, the
loci of the RGB and AGB are relatively clearly separated (see,
for example, \citealt{simoda70, buonanno81, sandquist96,
  sandquist04}), making it a particularly useful cluster for studying
AGB stars. \citet{zinn77} classified a number of AGB stars as having
very weak $G$-bands, and subsequently \citet{smith93} found that a
substantial fraction of AGB stars in M5 have enhanced CN band
strengths. The presence of CN-strong stars on the AGB of various
globular clusters has been reviewed by \citet{snedenAGB} and
\citet{campbell96}, based on the relatively sparse literature
available. Table 1 in \citet{campbell96} suggests that in clusters
more metal-poor than M5 the AGB stars tend to have weak CN
bands. \citet{campbell10} confirmed a relatively large population of
both CN-weak and CN-strong stars on the AGB of M5. This cluster
therefore offers an opportunity for a more extensive study of
N-Na-Mg-Al element enhancements on the AGB.

Two general mechanisms have been proposed to explain these abundance
patterns (e.g., \citealt{kraft94}). The first is that the surfaces of
these stars are polluted during the RGB phase by the interior products
of proton-capture reactions which have been consequently mixed to the
surface. The second is that the stars with high N, Na, and Al and low
O and C are part of a second generation of stars, formed out of gas
ejected by polluters in which hot H burning took place. The nature of
the polluters is still debated, including intermediate-mass AGB stars
from $\sim$4-8 $M_{\odot}$ \citep{ventura09}, fast-rotating massive
stars with 20-60 $M_{\odot}$ \citep{decressin07}, and massive binaries
with $\sim$ 20 $M_{\odot}$ \citep{demink09}. The discovery that these
abundance patterns continue to at least the main-sequence turnoff
indicates that these must be second-generation stars with the
abundance anomalies throughout the entire star.

However, the surface abundances of some elements and isotopes are
affected as stars go through the later phases of evolution
(e.g., \citealt{gratton00,sm03}).  In addition to the first dredge-up
on the low RGB, ``deep mixing'' or ``extra mixing'' on the upper
reaches of the RGB causes C and Li abundance drops, N increases, and
$^{12}$C/$^{13}$C ratio decreases which were not predicted in
original stellar models. This requires an additional physical effect
that has not been conclusively identified. Possibilities include
magnetic buoyancy (e.g. \citealt{busso07, denissenkov09}) and mean
molecular weight gradients which lead to ``thermohaline'' mixing
\citep{eggleton06, eggleton08, charbonnel07}. Part of the uncertainty
lies in the efficiency of the mixing by either mechanism. For example,
\citet{charbonnel07} showed that thermohaline mixing could account for
the abundance patterns on the RGB if the efficiency for mixing was
high, but two-dimensional simulations by \citet{denissenkov10} of thermohaline
mixing found that the actual efficiency was much smaller, closer to
the magnitude in \citet{kippenhahn80}.

There is also an ongoing debate about whether extra mixing happens on the
AGB. Models without AGB extra mixing may have difficulty explaining
observations of C/N and $^{12}$C/$^{13}$C ratios in AGB stars (e.g.,
\citealt{lambert86, lebzelter08, milam09}) and O isotope ratios in
pre-solar grains (e.g, \citealt{hoppe97}). \citet{karakas10} argued
that if extra mixing on the RGB was included in the models, then no
extra mixing on the AGB was needed to explain the C/N and C and O
isotope ratios dredged up to the surface and observed in stars, at
least at solar metallicities. \citet{busso10} found, however, that
extra mixing in AGB stars was necessary to match isotope ratios in
pre-solar grains, along with the C isotope ratios in C(N) stars, even
if extra-mixing was included on the first ascent RGB. However, the
mechanism for this extra mixing, like its counterpart on the RGB, is
not yet known.

If deep mixing can also occur in AGB stars, then there is the
possibility of CNO abundance differences being produced between the
RGB and AGB. Recently, stellar models have been evolved from the
main-sequence to the thermally pulsing AGB that includes mechanisms for
mixing and extra mixing to trace the evolution of surface abundances
in low-mass stars. \citet{stancliffe10} focused on low-metallicity
stars. Thermohaline efficiency was adopted from \citet{charbonnel07}
and is therefore very high. They found that $^3$He is not all
depleted, so thermohaline mixing could persist on the AGB. In
addition, on the early AGB, the deepening of the convective envelope
changes slightly the surface Li and $^3$He abundances and the C
isotope ratios.

In summary, theoretical work shows there are potentially interesting
changes in the light elements resulting from mixing and extra mixing throughout
the RGB and AGB. Observational evidence of the existence and size of
these effects will constrain the mechanism of mixing and its
efficiency. The AGB stars of M5 are cleanly separated from RGB stars,
and the stars at the tip of the RGB provide a good reference for
light-element abundances that may change on the AGB. In this paper we present
measurements of light-element abundances for stars on both the RGB and
AGB of M5 in an effort to determine whether there is any variation
with stellar evolution.

The star-to-star inhomogeneities in the light elements of M5, however, do not
appear to extend to the heavy elements formed by neutron-capture
processes. This suggests that their production is divorced from the
nucleosynthesis in the self-polluting cluster stars that made the
second generation stars. The first in-depth study of Ba, La, and Eu in M5 from \citet{ivans01}
found small internal scatter and good overall agreement with halo
field subdwarfs with similar [Fe/H]. Interestingly, \citet{ivans01}
noted that the abundances in M4 \citep{ivans99} showed
enhanced $s$-process contributions from AGB stars that enriched the
natal gas of all stars in the cluster. \citet{yong08b} measured 27
elements heavier than Fe in two RGB stars in M5 and 12 RGB stars in
M4. In addition to confirming the differences found by
\citet{ivans01}, they found that the abundance ratios could be
explained by some $s$-process in M5 as well, though at a much smaller
fraction than M4.

No freshly $s$-processed material is expected to appear on the surface
of the present-day M5 AGB stars, because third dredge-up does not
occur for stars with $M< 1.5M_{\odot}$. However, as is clear from the
discussion of extra mixing, we do not fully understand the
possible mixing events that can occur outside of the long-established
dredge-up events, and the $s$-process elements for stars
with a range of evolutionary states in M5 provide an opportunity to
test models. For example, \citet{masseron06} suggest that the
$s$-process enhancements seen in the extremely metal-poor AGB star CS
30322-023 are the result of an unknown mixing process that has brought
this just-produced material to the surface. In this study, we also explore
the origin of the neutron-capture elements in M5, and
if there are any signatures of an $s$-process contribution.

\section{Observation Details}

\begin{figure*}
\plotone{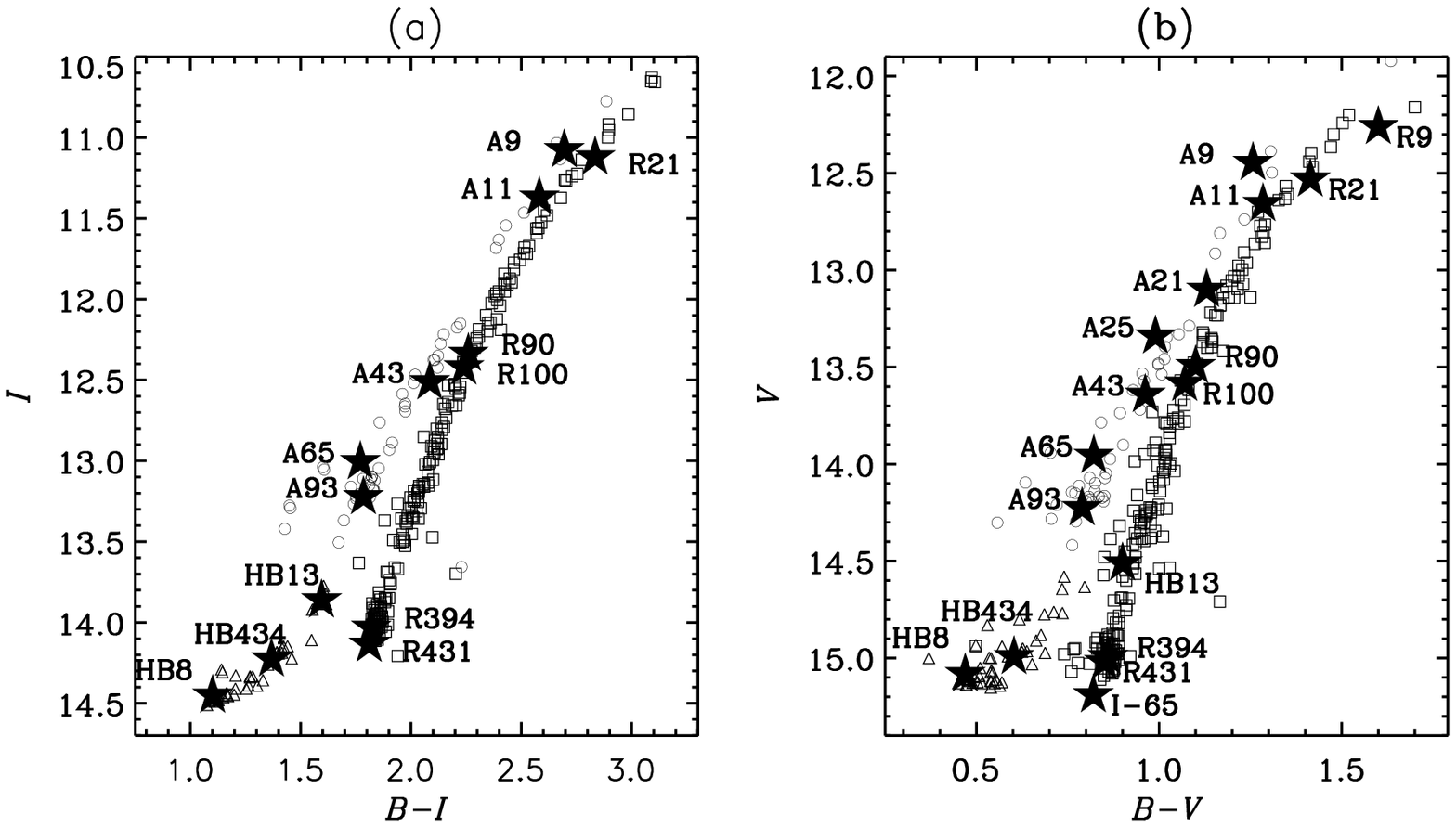}
\figcaption{
CMD of M5 in two different colors: (a) $B-I$
and (b) $B-V$. The circles correspond to AGB stars, the triangles to
HB stars, and the squares to RGB stars.  The stars included in our
HIRES observational sample are plotted with the filled star symbols.
\label{CMD}}
\end{figure*}

The stars observed for our investigation were chosen to sample the
RGB, the RHB, and the AGB of M5. Much of our sample is covered by the
photometric study of \citet{sandquist04}, which shows a clear
separation of the RGB from the AGB in their CMDs. The selection of targets was made so as to
have a relatively broad coverage of each evolutionary branch up to the
RGB tip luminosity, with a more uniform spread in the AGB to test for
any subtle evolutionary effects that might occur. Other than avoiding
the most crowded central regions, no other selection was used in
choosing targets. In Figure \ref{CMD}, we plot our sample within CMDs of
M5 obtained from the $BVI$ photometry of \citet{sandquist04}.

The observation details along with photometry are presented in Table
\ref{obsdetails}. All of the photometry data are taken from
\citet{sandquist04}, except for the RGB star I-65, which was taken
from \citet{buonanno81}. We have adopted a naming convention based on
the compilation of \citet{sandquist04}, where each star is designated
by either A, R, or H based on whether it is an AGB, RGB, or HB star,
respectively. The number following is based on the ordering in the
respective photometry tables in \citet{sandquist04}. However, one of
the designated HB stars, HB13, seems to be better characterized as an
AGB star (see Figure \ref{CMD} and Tables \ref{abund1} $-$
\ref{abund6}). For added clarity, we add our own classifications to
the last column of Table \ref{obsdetails}, along with designations
from \citet{arp1955} where available.

Our spectra were obtained using the HIRES spectrograph on the Keck 1
telescope \citep{vogt94}. On 2007 June 5$-$7 we observed the AGB, HB,
and part of our RGB sample with the recently upgraded detector focal
plane, a setup that provided near continuous wavelength coverage
between 3700 and 7950 \AA{}. The remainder of the RGB sample is from an
earlier run on 2000 June 6 with the original HIRES CCD, and therefore
covers a more limited wavelength range of 5300$-$7600 \AA{}.

\section{Stellar Parameters and Analysis}

\subsection{Reductions and Model Atmospheres}

We reduced our spectra using the MAKEE data reduction
package\footnote{http://spider.ipac.caltech.edu/staff/tab/makee/}. Equivalent
widths (EWs) were measured using the SPECTRE program \citep{spectre} by
fitting Gaussian profiles. For strong lines with extended wings,
direct integration of the line profile was used to measure the EW. 

\begin{figure}
\begin{center}
\scalebox{.6}[.6]{
\plotone{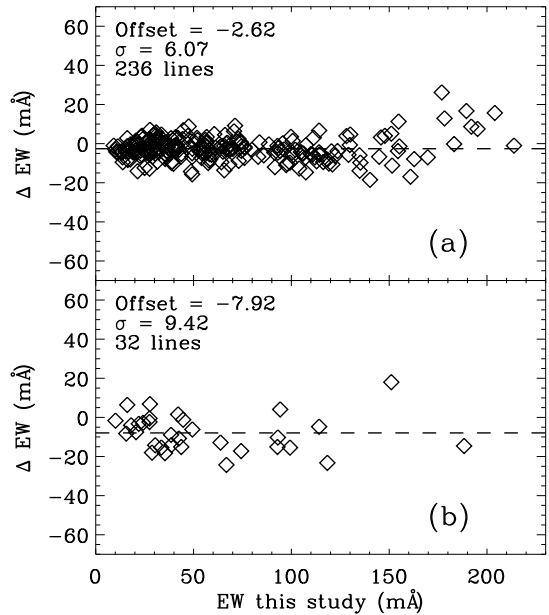}}
\end{center}
\figcaption{
(a) Comparison of EWs measured by our study and \citet{rc03}. Plotted
  are the differences between the EWs, in the sense of our study minus
  \citet{rc03}, vs. the EWs from our study.
(b) Comparison of EWs measured by our study and
  \citet{ivans01}. The EWs from \citet{ivans01} originally come from
  \citet{sneden92}, and have been transformed to their Keck scale. The
  difference $\Delta$EW is in the sense of our study minus
  \citet{ivans01}.
\label{ewrc03}}
\end{figure}

The line list was based partially on those used in \citet{ivans03} and
\citet{rc03} and, when available, the \gf{} values were updated with more
recent measurements. Many additional lines, predominantly in the blue
region, were also included. We present EWs and
atomic parameters with \gf{} references in Table \ref{EW}.  In Figure \ref{ewrc03} the
measured EWs are compared with those of \citet{ivans03} and
\citet{rc03} respectively. There is one star overlapping with both
studies, A11 (IV-59). In both cases we have on average slightly
smaller EWs, the consequences of which are discussed in Section
\ref{comparisons}.

We adopted the model atmospheres computed by \citet{kirby09}. These
were built on the ATLAS9 model atmospheres \citep{kurucz93}, using
updated opacity distribution functions \citep{castelli05}. In this grid
of atmospheres we set to [$\alpha$/Fe]$=0.3$ in all cases based on previous
studies \citep{ivans01,rc03}. In practice, the final abundance results
are fairly insensitive to variations in the adopted [$\alpha$/Fe]
ratio of the model atmosphere. We then interpolated the atmospheres to
the final \teff{}, \logg{}, and [Fe/H] values. The derivation of the
parameters is described in Section \ref{atmospheres}. The current
version of the local thermodynamic equilibrium (LTE) spectral analysis code
MOOG\footnote{http://verdi.as.utexas.edu/moog.html} \citep{sneden73}
was then used for the EW and spectral synthesis abundance determinations.

We accounted for hyperfine splitting (HFS) in the \ion{Sc}{2},
\ion{V}{1}, \ion{Mn}{1}, \ion{Co}{1}, and \ion{Cu}{1} lines with
the HFS parameters given by
Kurucz\footnote{http://kurucz.harvard.edu/linelists.html}. We used the
HFS parameters from \citet{mcwilliam98}, \citet{lawlerEu} and
\citet{lawlerLa} for the \ion{Ba}{2}, \ion{Eu}{2} and \ion{La}{2}
abundance determinations, respectively. The \ion{Th}{2} abundance was
determined from spectral synthesis of the 5989 \AA{} transition with
the \gf{} value from \citet{nilsson02}. Using the line parameters
from \citet{lucatello03}, the C abundance and \c12{} values were
determined from spectral synthesis of the CH $G$-band regions near 4234
and 4360 \AA{}, and N was determined from the CN band at 3880 \AA{}.

\subsection{Stellar Parameters \label{atmospheres}}

We took a hybrid approach  for deriving the stellar parameters. The
\teff{} value for each star was spectroscopically set by eliminating
any abundance trend determined from individual \ion{Fe}{1} lines with
their excitation potential (excluding lines with excitation potential
$\sim0.0$ eV). This, in practice, gives very good agreement to the
\teff{} calculated with the $B-V$ and $V-K$ color-\teff{} relationship
from \citet{ramirez05}, with $K$ magnitudes from the Two Micron All
Sky Survey (2MASS, \citealt{2mass}), and reddening of $E$($B-V$)$=0.03$ \citep{harris}. 
The microturbulent velocity ($v_t$) was set in the usual manner by
eliminating any trend of individual \ion{Fe}{1} line abundances with EW.

For the \logg{} determination we chose to use the known distance to M5
instead of ionization balance. This requires using additional
information, specifically the distance modulus, the stellar mass, and
the bolometric correction (BC) for each star. We adopt the distance
modulus given by \citet{kraft2003}, and following \citet{ivans01},
assume a mass of 0.80 $M_{\odot}$ for the RGB sample and 0.70
$M_{\odot}$ for our AGB and HB sample to account for expected mass
loss. This value may be slightly high, at least when comparing to mass
estimates from RRc Lyrae in M5. \citet{clement1997} and
\citet{kaluzny00} calculate average masses for their samples of RRc
Lyrae stars at 0.58 and 0.54 $M_{\odot}$, respectively. However,
assuming a mass of 0.60 $M_{\odot}$ generally decreases the \logg{} by
approximately only 0.07 dex, and therefore has little affect on the
abundance determinations. Given the small change to \logg{}, we use
the 0.70 $M_{\odot}$ value for consistency with \citet{ivans01}. We
then use the BCs from \citet{houdashelt}. Similar
to the technique described in \citet{rc03}, the tables of
\citet{houdashelt} are interpolated over [Fe/H], \teff{}, with a first
guess value of \logg{} to get an initial value of BC. This was then
used to recalculate \logg{}. The procedure was iterated until
self-consistent values were obtained for both BC and \logg{}.

Our method of deriving \logg{} using the known distance to M5 gives different values
than the traditional spectroscopic ionization balance technique
(Figure \ref{FeI-FeII}). This
is discussed in great detail for M5 in \citet{ivans01}, and more
generally for clusters in \citet{kraft2003}. In short, the possibility
of unaccounted for effects, e.g., non-LTE (NLTE) effects, will most greatly
affect the neutral \ion{Fe}{1} species \citep{thevenin}. Therefore 
the final adopted [Fe/H] for our model atmospheres is chosen to agree with
the final derived [\ion{Fe}{2}/H] value within 0.1 dex, instead of the
[\ion{Fe}{1}/H] value. Our final atmospheric parameters are given in
Table \ref{parameters}. The radial velocities of each star are also
presented in Table \ref{parameters}, with the error on each measurement
approximately 1 km s$^{-1}$ \citep{griest10}. 

Another important feature shown in Figure \ref{FeI-FeII} is the offset
between the Fe abundances derived for each evolutionary state. The AGB
stars show consistently lower \ion{Fe}{1} and \ion{Fe}{2} abundances
compared to the RGB stars. The RGB abundances are on average 0.16 dex
higher in \ion{Fe}{1} than for the AGB part of the sample. This is
almost the same as the 0.15 dex offset found by \citet{ivans01}. We
also find a similar offset in the \ion{Fe}{2} values of 0.13 dex in
the same direction (here \citealt{ivans01} find only a 0.06 dex
difference). The two HB stars also show a large difference in their
\ion{Fe}{1}$-$\ion{Fe}{2} values, along with an offset from the AGB and
RGB stars. 

We take the suggestion of our referee, and explore if the variations
in the number of \ion{Fe}{2} lines used in each star has an effect on
the offsets among the different evolutionary branches. First we define
a subset of \ion{Fe}{2} lines that are generally present in the stars
of the whole sample. This comes out to six lines, of which there is an
average of $\sim5$ measured per star. Interestingly, the AGB sample
then gives on average a $+0.04$ dex offset in [\ion{Fe}{2}/H] when
using only these lines, while the HB sample is offset by $+0.02$ dex,
and the RGB has no change on average. The difference in the line list
could account for some of the offset, but there remains an unexplained
significant difference in the values of [\ion{Fe}{2}/H]. These
differences seem to point to problems in our standard one-dimensional, LTE
analysis when dealing with these evolved stars. The caution most
strongly applies to absolute abundances, and is minimized somewhat by
considering abundance ratios. This is discussed further in Section
\ref{results}.

\subsection{Calculation of Abundance Errors}

We follow the abundance-error analysis technique described in
\citet{johnson2002}, which includes the dependences among
\teff{}, \logg{}, and $v_t$. To estimate the component of the error arising
from EW measurements and uncertainties in atomic parameters, we used
the standard error based on the standard deviation of the abundances
derived from multiple lines. For abundances with four or fewer
individual line measurements, we assumed a conservative 0.15 dex lower
limit for the standard deviation. 

For the atmospheric parameters, we assume errors of 100\,K in \teff{},
0.2 dex in \logg{}, and 0.2 km s$^{-1}$ in $v_t$. The \teff{} and
$v_t$ errors are motivated by the sensitivity of the \ion{Fe}{1} line
abundances. Changes of the order of 100 K and 0.2 km s$^{-1}$ and
larger in \teff{} and $v_t$, begin to introduce large trends of
\ion{Fe}{1} with excitation potential and EW, respectively. For the
\logg{} error, we follow the calculations of \citet{rc03}, which
combine uncertainty in distance, stellar mass, and \teff{}, to arrive
at the value of 0.2 dex.

We then calculated the abundance errors due to atmospheric
uncertainties using three representative stars, a low-\teff{} star (to
be used for stars with \teff{} $< 4500$ K), a moderate-\teff{} star
(for the temperature range $4500 \leq $\teff{} $< 5000$ K), and a
high-\teff{} star (for \teff{} $\geq 5000$K, which applies only to the
two hottest HB stars). These errors were then applied to each star in
the \teff{} ranges noted above The final combined errors are then
calculated using Equations (5) and (6) in \citet{johnson2002}.

\section{Results}

Our abundance results are summarized in Tables
\ref{abund1}$-$\ref{abund6}. In Table \ref{sigma} we give the average
values, the standard deviation $\sigma$, and number of measurements
for each abundance ratio for both the entire sample and for each
evolutionary branch.  We take the solar photospheric abundances from
\citet{anders89}, but for Fe we use solar $\log \epsilon ({\rm Fe}) =
7.52$. We note that for recent compilations of the solar abundances
(e.g., \citealt{asplund09, lodders09}) a three-dimensional analysis gives C, N, and O
abundances appreciably different from \citet{anders89}. However,
because we are performing a one-dimensional analysis, we use the \citet{anders89}
results. The neutral species ratios are reported relative to
\ion{Fe}{1} and the ionized species ratios are reported relative to
\ion{Fe}{2}. All abundances reported without an ionization state are
neutral, and are explicitly labeled as such when both neutral and
ionized states are measured for a given element.

The one exception to this is [O/Fe]. With their high excitation
potentials, the abundance determined from the O triplet lines at 7770
\AA{} are taken relative to the \ion{Fe}{2} abundance. The value
determined from the forbidden O lines at 6300 and 6363 \AA{} are still
expressed relative to \ion{Fe}{1}. When lines from both sets of
transitions are measured, these relative values are combined to obtain
the final [O/Fe] given in Table \ref{abund1}. Shown in Figure
\ref{oxygen} are the [O/Fe] abundances derived from the triplet lines and
the forbidden lines in the six stars where both sets of lines are measured.
In these cases, we find good agreement between both
abundance ratio determinations when using this method. 

We also apply NLTE corrections for abundances determined from the O
triplet lines and the Na lines. These corrections were applied
line by line by extrapolating the tables from \citet{gratton99}
between EW, metallicity, \teff{}, and \logg{}. It is important to note
that the NLTE corrections for Na from \citet{gratton99} can be in the
opposite direction from other studies (see the review by
\citet{asplund05}). However, we adopt the \citet{gratton99} study
because of the ability to extrapolate the NLTE corrections over a wide
range of EW, \teff{}, and \logg{} values, which is vital to this
study. Since we consistently use the \citet{gratton99} corrections, this
should leave the relative abundance trends in this study intact.
However, this may lead to systematic differences with other studies
that use different sets of NLTE corrections for Na (e.g.,
\citealt{nanlte,mashonkina00,shi04}).

\subsection{Comparison to Previous Studies \label{comparisons}}

We compare our atmospheric parameters and abundance measurements 
for the star A11 to those determined by \citet{ivans01} and \citet{rc03}
in Table \ref{comparison}. The results of \citet{rc03} are converted
to be relative to the \citet{anders89} solar values to be consistent
with our abundance ratios. The atmospheric values in the different studies 
are in good agreement. We find lower [\ion{Fe}{1}/H] and
[\ion{Fe}{2}/H] values, which can at least be partially explained
by our slightly smaller EW values. However, when we compare abundance
ratios we find generally good agreement with these previous
studies. The exceptions to this seem to be [Si/Fe], [\ion{Sc}{2}/Fe],
and [Cu/Fe].

To test the effects of the EW offsets we ran two sets 
of abundance determinations with our measured
EWs increased by 7.92 and 2.62 m\AA{}, to match the offsets found
between our study and \citet{ivans01} and \citet{rc03},
respectively. In both cases all abundance ratios relative to Fe
remained virtually unchanged. However, the [\ion{Fe}{1}/H] and
[\ion{Fe}{2}/H] abundances did experience shifts. For the
\citet{ivans01} EW shift we calculate [\ion{Fe}{1}/H]$=-1.37$ and
[\ion{Fe}{2}/H]$=-1.27$, in excellent agreement with their
values of $-1.40$ and $-1.25$. For the \citet{rc03} EW shift we
calculate [\ion{Fe}{1}/H]$=-1.51$ and [\ion{Fe}{2}/H]$=-1.41$, 
which within the errors is in agreement with their values of $-1.40$ and $-1.35$.

The discrepancies still remain for [Si/Fe], [\ion{Sc}{2}/Fe], and
[Cu/Fe]. The differences in abundance ratios are unlikely
to be due to atmospheric parameters given the good agreement 
between those adopted in the various studies and the 
insensitivity of abundance ratios to the correspondingly small differences 
(see, for example, Table 3 in \citealt{ivans01}). 

One possibility lies in the different sets of lines and/or log $gf$
values adopted here versus the previous studies. We test this by
deriving abundances based only on lines that overlap with \citet{rc03}
along with their log $gf$ values. We do not perform this same
exercise with \citet{ivans01} because there are only 32 overlapping
lines (versus 236 lines from \citealt{rc03}), making the comparison
less useful. The only significant deviation comes in the
[\ion{Cr}{1}/Fe] abundance ratio, which we find to be 0.24 dex higher
when using only overlapping lines. This is not due to log $gf$ values,
as we have adopted the same values from \citet{rc03}. The difference
comes in the small number of overlap lines in \ion{Cr}{1}, only three
in this case as compared to the 24 lines we use in the final analysis
of this star. These three lines give a relatively large scatter using
either set of EWs. With the exception of [\ion{Cr}{1}/Fe], however,
the abundance results that can be measured with this overlapping
subset of lines remain the same within errors.

To test the effect of using different atmosphere models, we ran an
abundance analysis on our A11 data using a MARCS model atmosphere with
an alpha enhancement of 0.4 \citep{marcs08}.  The results of this
exercise are also presented in Table \ref{comparison}. The only
atmospheric parameter that needed an adjustment was $v_t$. Overall
there is little change between the abundances derived from
our adopted atmosphere or the MARCS model atmosphere, with any of the
differences well within the errors. The lower [\ion{Fe}{1}/H] and
[\ion{Fe}{2}/H] values remain, and the [Si/Fe], [\ion{Sc}{2}/Fe], and
[Cu/Fe] values also remain very similar. Thus the sources of the
differences in these three particular [X/Fe] ratios between the
present work and the studies of \citet{ivans01} and \citet{rc03}
remain unidentified.

\section{Discussion}

\subsection{Abundance Comparisons \label{results}}

If there are no systematic errors in our analysis and if 
no changes in the surface abundances with evolutionary phase,
then we should see the same patterns when we compare the RGB, HB and
AGBs and when we compare our results with the literature values. However,
there are several cautionary points that need to be made before such
comparisons are attempted.

\subsubsection{Absolute Abundances \label{offsets}}

\begin{figure}
\plotone{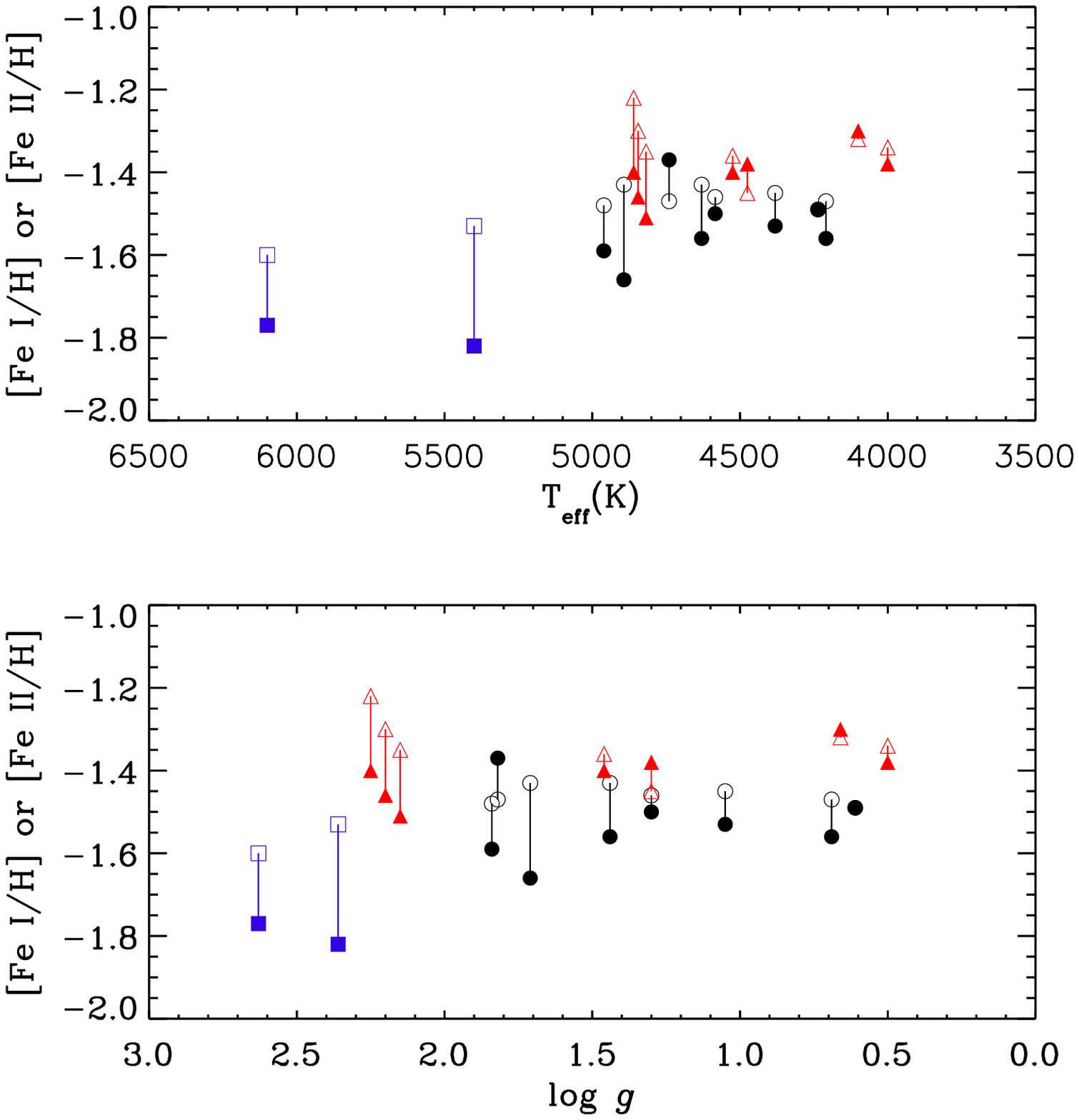}
\figcaption{
Our measurements of [\ion{Fe}{1}/H] (solid symbols) and
[\ion{Fe}{2}/H] (hollow symbols) vs. effective temperature and
surface gravity.  The circles correspond to AGB stars, the triangles
to RGB stars, and the squares to HB stars (colored black, red, and
blue, respectively in the electronic edition). The lines connect the
[\ion{Fe}{1}/H] and [\ion{Fe}{2}/H] measurements from the same
star. In most of the stars, [\ion{Fe}{1}/H] is lower than
[\ion{Fe}{2}/H]. There is also an offset of both the [\ion{Fe}{1}/H]
and [\ion{Fe}{2}/H] between different evolutionary states. 
\label{FeI-FeII}}
\end{figure}

\begin{figure}
\plotone{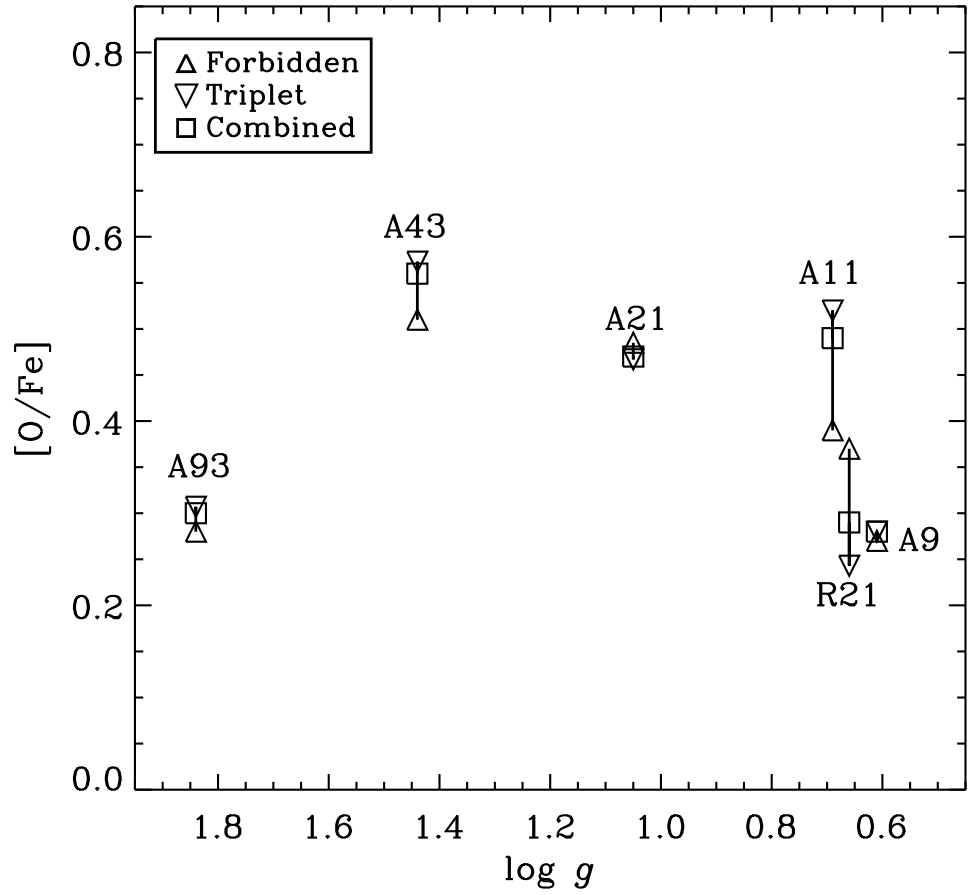}
\figcaption{
In six of our stars both forbidden and triplet lines of O were
measured. Plotted are [O/Fe] derived individually from both sets
of lines, along with the final combined abundance value 
vs. stellar surface gravity.
A solid line connects the measurements for each individual star. Note that the combined
[O/Fe] is not the average of the forbidden and triplet values because
there are different number of lines coming from each set (see Table \ref{EW}).
\label{oxygen}}
\end{figure}

As can be seen in Figure \ref{FeI-FeII}, there are some
difficulties in interpreting certain absolute abundances from our
analysis. Most notably, the Fe abundance appears to change with
evolutionary state. \citet{korn07} find that atomic
diffusion could operate on unevolved stars near the turnoff and hence
modify the surface abundance of elements such as iron. However, when
looking at stars evolved past the lower RGB, this mechanism only seems
to affect hotter BHB stars (\teff{}$>8500$K, e.g., \citealt{behr00}).
For our sample of cooler and evolved stars no internal processes are
known to modify the surface iron abundance of globular cluster
stars, which leads us to suggest that the trend in Figure
\ref{FeI-FeII} is artificial. Thus, comparisons between the absolute
abundances of other elements between the RGB, AGB, and HB stars may
also contain unaccounted for systematic effects.

There are some ways to minimize this problem. One method, when
available, is to compare abundance values that are less prone to
effects that potentially have significant repercussions on the
analysis, such as three-dimensional atmosphere and NLTE effects (e.g.,
\citealt{asplund05, asplund09}). For example, in Figure \ref{FeI-FeII},
[\ion{Fe}{2}/H] shows a far more consistent value than
[\ion{Fe}{1}/H]. This is probably because in the stellar
conditions of the stars we are analyzing [\ion{Fe}{2}/H] is
the majority species, and therefore less prone to showing NLTE
effects that we cannot account for. Another approach to
the problem is to use pre-determined corrections for NLTE,
as we have done for O and Na.

These concerns raise the issue of how to best compare the metallicity
derived here for M5 to results of previous studies. Given that the
majority of stars in prior studies are not evolved past the RGB phase,
the most meaningful comparison may be to only consider RGB stars in
our sample (note, though, that the direct star-to-star comparison
discussed in Section \ref{comparisons} is for an AGB star). We also
recommend only comparing [\ion{Fe}{2}/H] abundances instead of
[\ion{Fe}{1}/H] for the reasons given above. As reported in Table
\ref{sigma}, we find for our RGB sample an average
[\ion{Fe}{2}/H]$=-1.33$. This compares well to recent
high-resolution studies giving [\ion{Fe}{2}/H] values of $-1.27$
\citep{koch10}, $-1.32$ \citep{carretta09b}, $-1.33$ \citep{rc03},
and $-1.20$ \citep{ivans01}. All of these values have been scaled
to our adopted solar abundance of $\log \epsilon({\rm Fe}) =
7.52$, and only include stars that have not evolved past the RGB.

\subsubsection{Abundance ratios}

\begin{figure}
\begin{center}
\scalebox{.5}[.5]{
\plotone{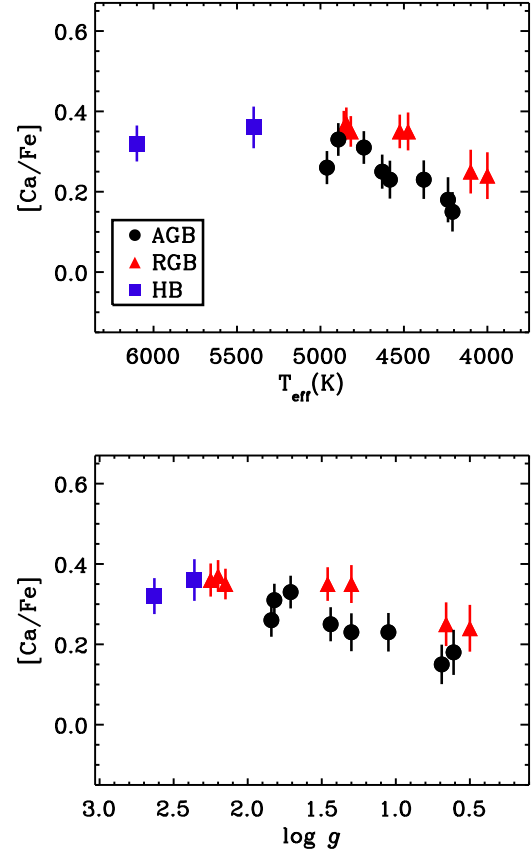}
}
\end{center}
\figcaption{
Our values of [Ca/Fe] plotted against \logg{} and \teff{}. There is an offset in [Ca/Fe] of $\sim0.08$
dex between the AGB and RGB stars. 
\label{ca}}
\end{figure}

\begin{figure}
\begin{center}
\scalebox{.5}[.5]{
\plotone{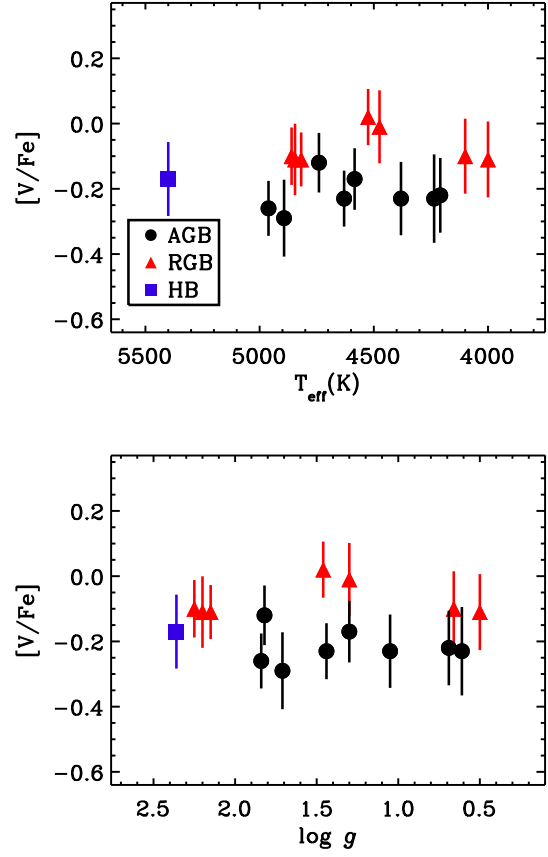}
}
\end{center}
\figcaption{
Our values of [V/Fe] plotted against \logg{} and \teff{}. Like with
[Ca/Fe], there is an offset between the AGB and RGB stars, but in this
case it is 0.15 dex. 
\label{V}}
\end{figure}

\begin{figure}
\begin{center}
\scalebox{.5}[.5]{
\plotone{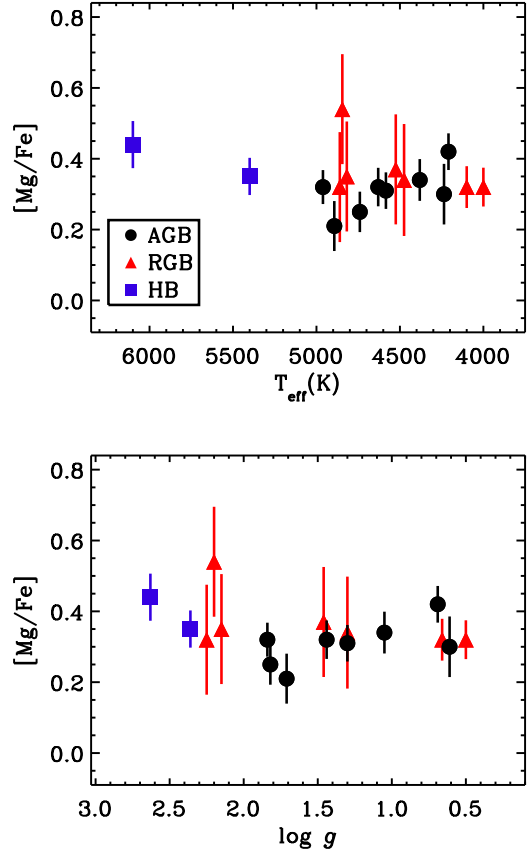}
}
\end{center}
\figcaption{
Our values of [Mg/Fe] plotted against \logg{} and \teff{}. Within the errors, all of the stars are consistent with a
constant [Mg/Fe] value of 0.34. 
\label{mg}}
\end{figure}

To minimize the effects of absolute abundance offsets, we now limit
our discussion to comparisons of abundance ratios, and discuss
element abundances relative to either \ion{Fe}{1} or \ion{Fe}{2}. This
assumes that the same NLTE, three-dimensional, and other issues affect the abundance
determinations of individual elements in a similar manner. However,
this still leaves some uncertainty, as illustrated in Figures \ref{ca},
\ref{V}, and \ref{mg}. For the abundance ratio of [Ca/Fe] shown in
Figure \ref{ca}, we find an offset depending on a stars evolutionary
state.  An offset is even more apparent for [V/Fe], shown in Figure
\ref{V}. In the case of [Mg/Fe], an abundance ratio that has been
shown to vary in globular clusters, Figure \ref{mg} shows that our
values are consistent within the errors with a constant value across
all observed stars (and also consistent with \citealt{carretta09b} who
found no Mg variation in a sample of 14 M5 RGB stars).

The abundances of Ca and V, like Fe, can only be modified by previous
stellar generations via supernovae ejecta, and therefore should not
vary as a star evolves. While Mg can be modified by the Mg$-$Al cycle,
the direct evidence of this is generally difficult to detect given the large
absolute amount of Mg in a star relative to the absolute amount of Al
present. This makes any potential systematic errors in the [Mg/Fe]
ratio easily able to mask true signs of variation. For [Ca/Fe] and
[V/Fe], which we expect should be constant, it seems that there are
systematic errors that give the appearance of
abundance differences between AGB and RGB stars.  The offset is 0.15
dex in the case of [V/Fe]. This is the largest abundance offset
between the AGB and RGB for elements heavier than Al and with greater
than one measurement on each branch (see Table \ref{sigma}). We can
take this as a conservative lower limit to our true abundance ratio
uncertainty when comparing stars from different evolutionary
states. A possible method to reduce these effects is to use
the differential analysis as described in \citet{koch08b,koch10}.
However the wide range of effective temperatures in our sample would both make this
type of analysis impractical and still make star-to-star comparisons problematic.

The potential for such systematic effects has ramifications beyond
comparing the different stellar evolution branches of M5. A recent
example comes from efforts to ascertain whether Ca varies within
globular clusters, and the implications for the origin of light
element variations found in these systems
\citep{leenature}. \citet{carretta10a} find from their large sample of
globular cluster red giant stars that Ca does not vary significantly
in a given cluster. M5 has a fairly cleanly separated AGB, unlike some
other globular clusters. Therefore, even though \citet{carretta09c}
show that AGB star contamination is low in their sample, any AGB
contamination could artificially inflate star-to-star Ca
variations. This effect would tend in the direction of making their
findings of already low Ca variations an upper limit.

\subsection{Abundance Inhomogeneities among the Lighter Elements Li
  through Mg}

Given the caveats above, we do find true abundance variations in the
lighter elements of C through Al among the stars in our M5
sample. Interesting changes in C, N, Li and the C isotopes may
potentially occur between the tip of the RGB and the onset of the
thermally pulsing AGB depending on (1) the existence and strength of
extra mixing and (2) the depth from which any of the material with
O$-$Na$-$Al abundance anomalies originates. Detecting these changes in M5
is complicated by the dispersion in these abundances for stars in all
evolutionary states. In the following sections, we explore if there is
evidence for any additional mixing or changes on the AGB branch that
manifest themselves in surface abundances, and how our sample fits
into the context of previous studies of M5.

\subsubsection{C, N, O and $^{12}$C/$^{13}$C among the Different Branches}

\begin{figure}
\begin{center}
\scalebox{.35}[.35]{
\plotone{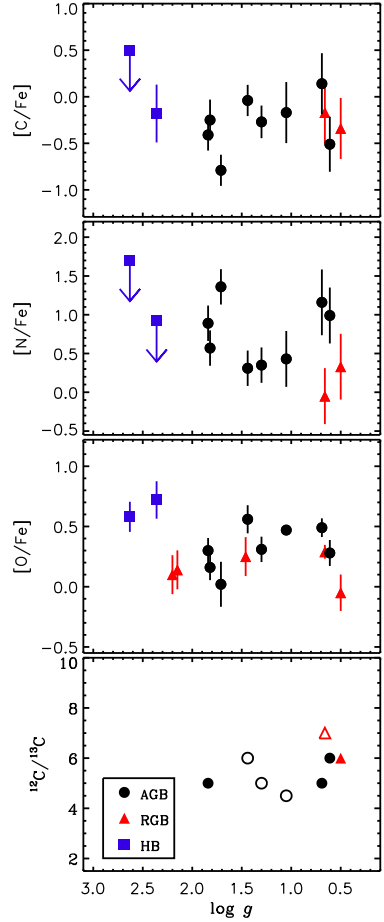}
}
\end{center}
\figcaption{
[C/Fe], [N/Fe], [O/Fe], and $^{12}$C/$^{13}$C vs. surface
gravity for stars in our M5 sample. There does appear to be some
variation of the derived [C/Fe] abundance ratios. However with the
large uncertainties, all but one of our stars is consistent with
having [C/Fe]$\sim-0.2$.  Among the AGB stars there is a notable
variation in [N/Fe]. The comparison between the RGB stars and AGB
stars for [N/Fe] in our sample is hindered by the small number of the
former, and possible systematic differences in the atmospheres of
these two evolutionary groups. [O/Fe] shows star-to-star variations,
although not as large as with [N/Fe]. For the $^{12}$C/$^{13}$C plot,
the filled symbols represent stars with [Na/Fe]$\ge0.15$, while the
hollow symbols represent stars with [Na/Fe]$<0.15$.  While only able
to measure the \c12{} ratio for a handful of stars, we find that the
results are all consistent with a value of $\sim5$. The error bars on
all of the measurement are $\pm2$. In particular, we find the same
\c12{} ratio in the RGB stars as in the AGB stars of our sample and
that there is no difference in values in the Na-high vs. Na-low
stars.
\label{cnoc12}}
\end{figure}

The light elements C, N, and O show variations that do not seem to be
correlated with either evolution along a particular evolutionary branch
(for which surface gravity is a proxy) or among different evolutionary
branches as a whole (Figure \ref{cnoc12}). Thus we do
not see any evidence for additional mixing within AGB stars compared
to the upper RGB stars in our sample. The HB stars have the highest measured [O/Fe]
abundances, but given the differences in effective
temperature, and to a lesser extent surface gravity, between the HB
stars and those of the RGB and AGB, a claim of a truly intrinsic
[O/Fe] enhancement in the HB stars is not made here.

Perhaps a more definitive test for evolutionary differences between
AGB and RGB stars is provided by the measurement of the
$^{12}$C/$^{13}$C ratio, which is less sensitive to model atmosphere
uncertainties. This ratio can be altered by the CNO cycle of H-burning
reactions. In Figure \ref{cnoc12} the $^{12}$C/$^{13}$C ratio is also
plotted versus surface gravity for the AGB stars and the two RGB stars
for which it could be measured. The values are all consistent within
the errors to be around 5.5, close to the CNO cycle equilibrium value
of 3.5. There is no difference between the RGB and the AGB stars, and
no change with AGB evolution. In particular, the low C isotope ratios
measured here are consistent with values found for three red giants in
M5 by \citet{pavlenko03}, as well as with values found on the upper
RGBs of other globular clusters and in halo field red giants (e.g.,
\citealt{brown89, gratton00, keller01, pavlenko03, shetrone03b,
  rb07}).

In the context of the AGB models of \citet{stancliffe10}, we
do not see any drop in the $^{12}$C/$^{13}$C going up the AGB as might
be expected from the thermohaline mixing assumed in their
models. However, AGB stars in M5 are at a higher metallicity and are
slightly lower mass than the most appropriate model of
\citet{stancliffe10}, so an exact comparison cannot be done. Also, it
appears that the AGB stars of M5 start with a much lower
$^{12}$C/$^{13}$C ratio than their models, which could make surface
modifications to this ratio difficult to detect.

\subsubsection{Evidence of the CNO cycle}

\begin{figure}
\begin{center}
\scalebox{.5}[.5]{ 
\plotone{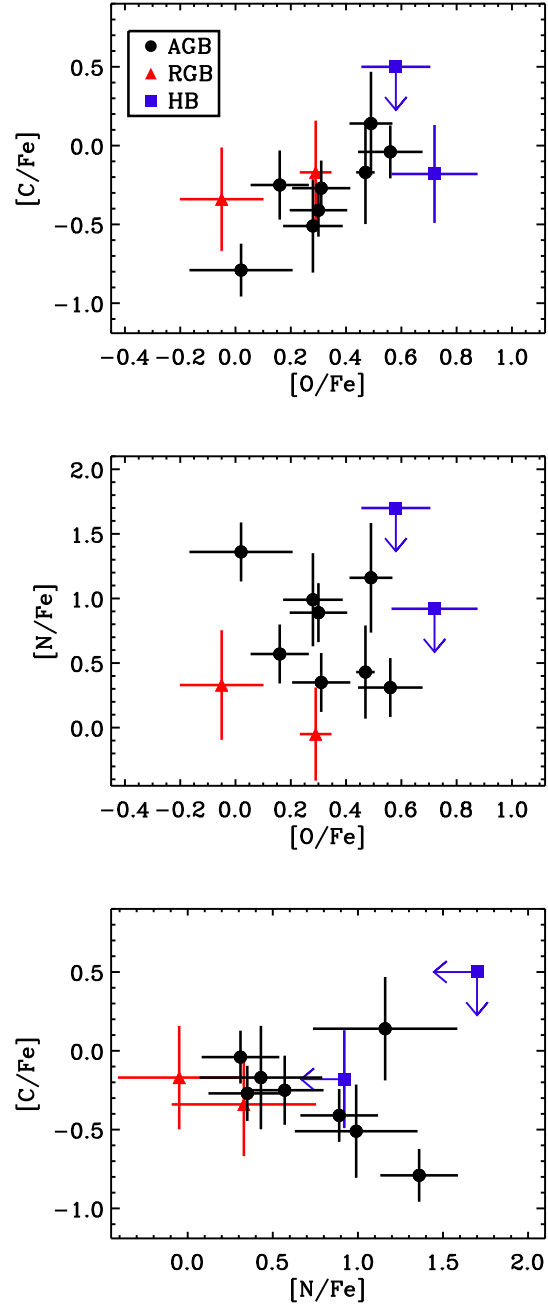} 
}
\end{center}
\figcaption{ Relative behavior of the CNO abundances among the M5
  sample. At least among the AGB stars, there appears a clear C$-$O
  correlation and C$-$N anticorrelation, although evidence for an O$-$N
  anticorrelation appears less secure. Such trends are signatures of
  the CNO cycle having affected the abundances of these stars.
\label{cno}}
\end{figure}

With only two RGB stars with [C/Fe] and [N/Fe] measurements, along
with possible systematic differences in atmospheres between RGB and
AGB stars, we generally limit the comparison of the CNO behavior
within the sample of AGB stars. The C$-$O correlation and C$-$N
anticorrelation expected from varying degrees of CNO cycling of
material appear to be present (Figure \ref{cno}). The scatter in the N
versus O plot is large enough to preclude any conclusions as to the
presence of an N$-$O anticorrelation.

\begin{figure}
\begin{center}
\scalebox{.5}[.5]{
\plotone{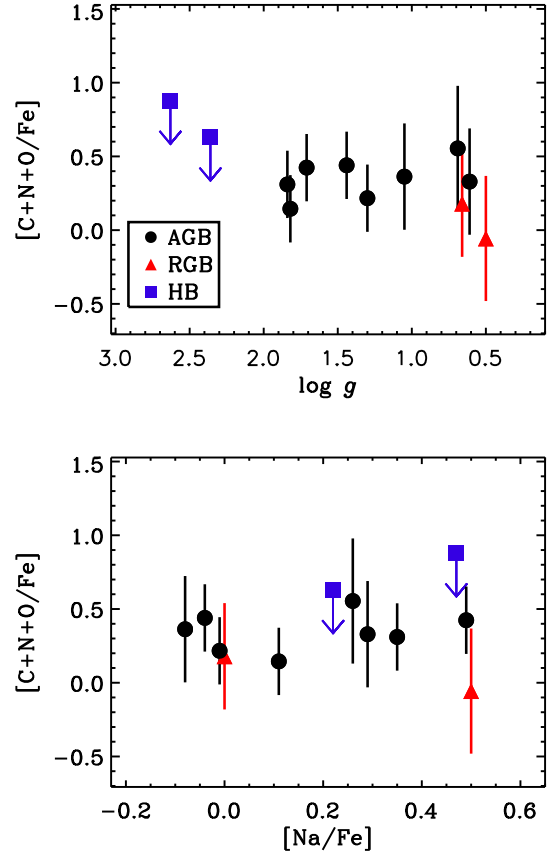}
}
\end{center}
\figcaption{
[C+N+O/Fe] values are consistent with being constant for the stars
in which it could be measured. The average for these stars is $\langle
{\rm [C+N+O/Fe]} \rangle = 0.37$.
\label{cpnpo}}
\end{figure}

We also see constant C+N+O in our sample, consistent with the material
in these stars having undergone CNO cycling in the first generation of
stars, in early RGB phases, or both. In Figure \ref{cpnpo}, we show
that the combined [C+N+O/Fe] ratio depends on neither evolution nor other
light-element variations. Combined with the lack of correlation of C,
N, O, and $^{12}$C/$^{13}$C with evolutionary state, this is an
indication that while the CNO cycling of material occurred, we do not
find evidence for extra mixing having occurred between the RGB tip and
the TP$-$AGB phase of M5 in any appreciable manner.

These results for the AGB are analogous to those found on the subgiant branch by
\citet{cbs02} and the RGB by \citet{sneden92} and
\citet{ivans01}. The presence of N-enhanced stars extending from the subgiant
branch to the AGB implies that they contain material that has been
processed through the CNO cycle of hydrogen burning prior to their formation.

\subsubsection{[C/Fe] and [N/Fe] as Compared to Previous Studies}

The wavelength coverage for the majority of the RGB sample precluded
measurement of [C/Fe]; however, we find the average [C/Fe] of the two
RGB stars we could measure, R9 and R21, is found to be $-0.25$ dex. Previous
studies of [C/Fe] in M5 have found values on the order of
0.2$-$0.4 dex lower, albeit using different techniques (\citealt{langer85} and
\citealt{smith97} from low-resolution spectroscopy of the $G$ band, and
by \citealt{pavlenko03} from the 2.3 $\mu$m CO bands).

\begin{figure}
\begin{center}
\scalebox{.5}[.5]{
\plotone{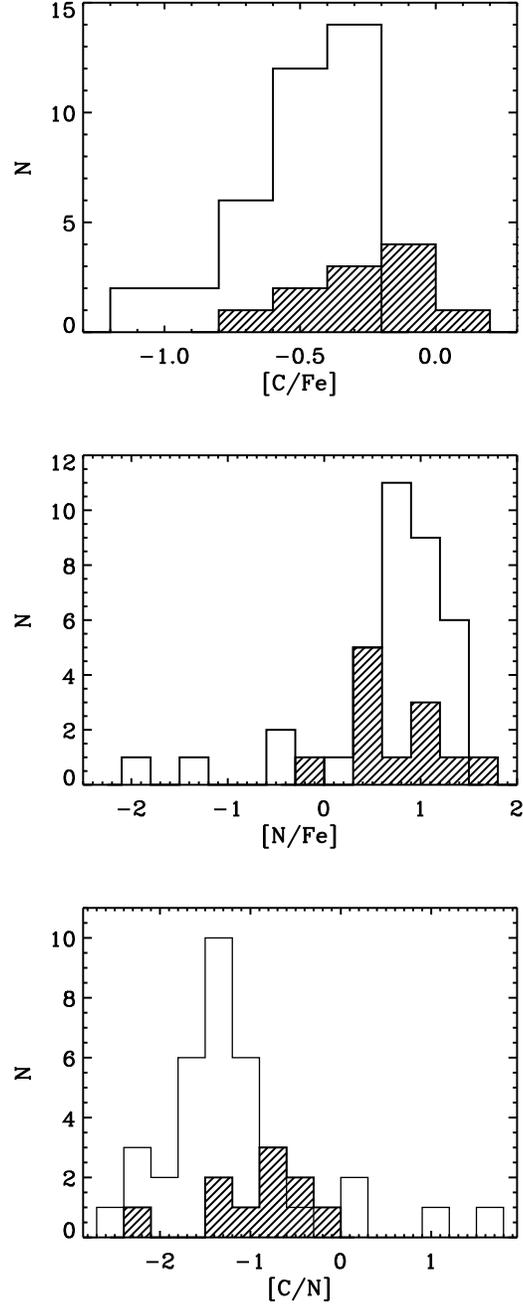}
}
\end{center}
\figcaption{
We show our [C/Fe], [N/Fe], and [C/N] distribution as compared to the
study of \citet{cbs02} for the same quantities. The filled histograms
represent this work, and the empty histograms \citet{cbs02}. For [C/Fe]
we choose a binning of 0.2 dex, while the binning is 0.3 dex for
[N/Fe] and [C/N] due to the larger distribution of values in these
latter two quantities. 
\label{cn}}
\end{figure}

We can further compare our entire sample of [C/Fe] and [N/Fe]
measurements to previous studies of M5. In particular, we use the work
of \citet{cbs02}, because of their large sample size. In Figure
\ref{cn}, we compare the distributions of [C/Fe], [N/Fe], and
    [C/N]. Our sample is more [C/Fe]-enhanced and slightly
    [N/Fe] deficient relative to that of \citet{cbs02}. This translates to
    an enhancement of [C/N] in our sample. If anything, because the
    stars from \citet{cbs02} fall below the RGB bump, one would expect
    from deep mixing that our sample would have lower [C/Fe] and
    higher [N/Fe] distributions. 

One possible explanation for these discrepancies is systematic offsets
of CNO abundance ratios between the two giant branches as discussed in
Section \ref{offsets}. However, except for [N/Fe], the RGB sample shows
similar distribution of values in [C/Fe] and [O/Fe] as the AGB
sample. Another possible explanation is a systematic difference
between our high-resolution abundance analysis and previous work using
CH and CN indices such as done in \citet{cbs02}.

Although systematic errors cannot be ruled out, one
intriguing possibility is that this discrepancy is related to the
finding of \citet{campbell10} that there is a relatively large number
of CN-weak stars on the AGB of M5 relative to the CN-weak stars on the
RGB. Given that \citet{cbs02} is a study of subgiant stars,
and our sample is predominately AGB stars, our higher peaked [C/N]
distribution would be expected from the \citet{campbell10}
result. Physically, this may be related to
high-N stars on the RGB having increased mass-loss rates. This leads
these stars to evolve to the bluer end of the HB, and possibly never
making it to the AGB \citep{norris81,campbell10}.

\subsubsection{Li}

We are only able to measure Li in three of our
stars, using spectral synthesis of the $\lambda6708$ Li resonance
doublet. In the red giants R394, I-65, and R431, which sit right at
the RGB bump at $V=14.99$, we find log$\epsilon({\rm Li})=0.76,$
0.73, and 0.96. For all other stars in our sample, Li could not be
detected and we find upper limits of
log$\epsilon({\rm Li})<0$. These log$\epsilon({\rm Li})$ values are
intermediate to the values found by \citet{gratton00} for field RGB
stars of similar metallicities below the RGB bump
(log$\epsilon({\rm Li})\simeq 1.15$) and above it
(log$\epsilon({\rm Li})< 0$). They are also consistent with the
log$\epsilon({\rm Li})$ values found on the RGB bump of M4 found by
\citet{dorazi10}. 

These measurements support the notion that deep mixing is performing
as expected from these previous studies on the RGB of M5. Unfortunately,
because we have only three stars at the RGB bump, we cannot comment on
the potential difference in the evolution of log$\epsilon({\rm Li})$
value for Na-rich and Na-poor stars as \citet{dorazi10} find in
M4. However, our consistent finding with the field and M4 suggests M5
as being a promising cluster for a similar study.

\subsubsection{Na, Al, and Mg}

\begin{figure}
\begin{center}
\scalebox{.8}[.8]{ 
\plotone{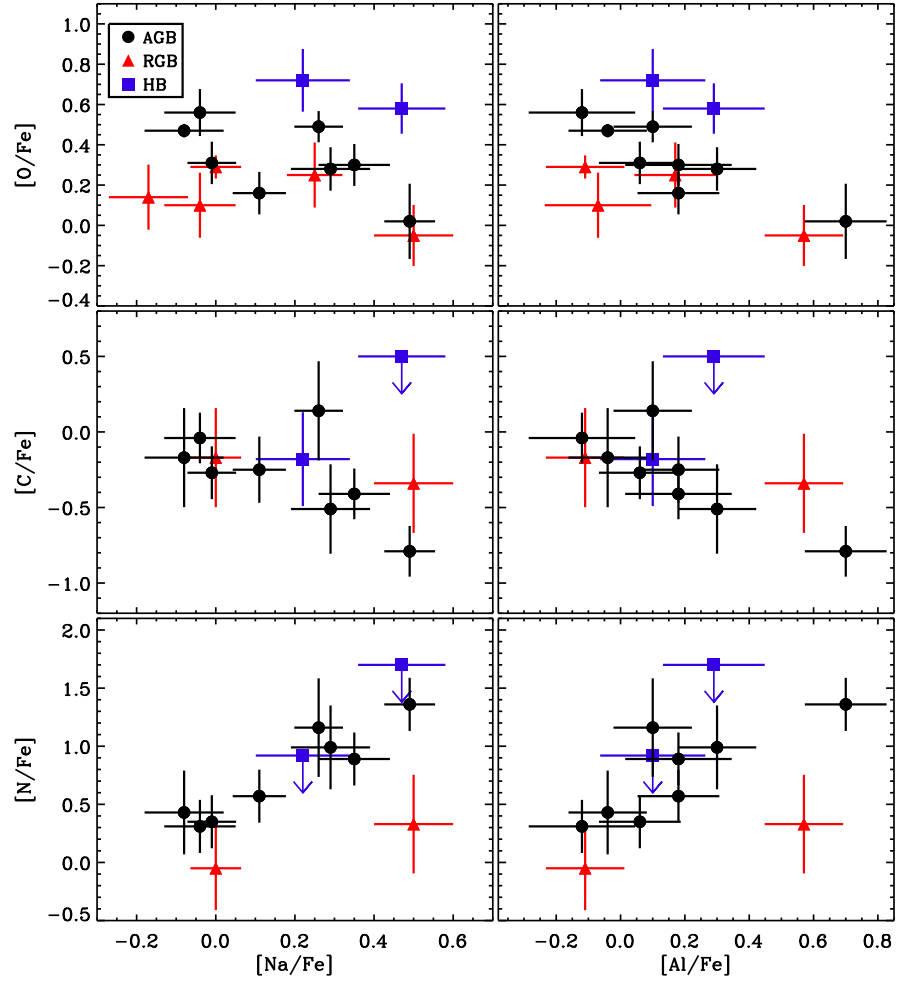} 
}
\end{center}
\figcaption{ CNO abundances are shown plotted vs. [Na/Fe] and
  [Al/Fe].  Our data are consistent with an O$-$Na and C$-$Na
  anticorrelation among the AGB stars, but inconclusive for the RGB
  and RHB stars.  However, there is a reasonably strong N$-$Na
  correlation, with the caution that the data for the two RGB stars
  may be systematically offset from that for the AGB stars. Similar
  (anti)correlations are apparent for O$-$Al, C$-$Al, and N$-$Al. We note
  that for two of our stars, HB434 and HB8, [Al/Fe] could only be
  derived from using the Al resonance lines at 3944 and 3962
  \AA{}. For these two stars, an NLTE correction is adopted from
  \citet{alnlte} of +0.65 dex.
\label{NaAlN}}
\end{figure}

\begin{figure}
\begin{center}
\scalebox{.5}[.5]{
\plotone{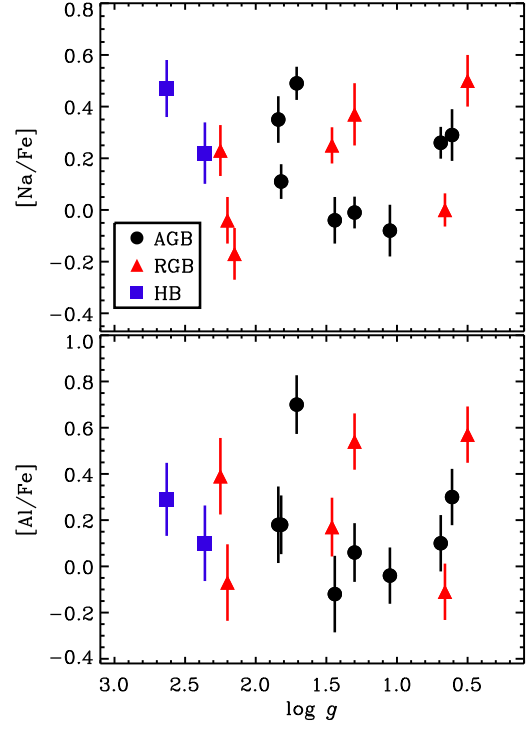}
}
\end{center}
\figcaption{
[Na/Fe] and [Al/Fe] abundances, shown here plotted vs. surface gravity,
show no discernible trend with evolutionary state or atmospheric parameters.
\label{NaAltrend}}
\end{figure}

The C, N, O, and Na abundances show the expected (anti)correlations if
the CNO cycle took place in the same location as the Ne$-$Na cycle
(Figure \ref{NaAlN}). In the same figure we also show the C, N, and O
abundances as a function of [Al/Fe]. Similar correlations as with
[Na/Fe] are indicative of the Mg$-$Al cycle, which we discuss more
below. In Figure \ref{NaAltrend}, both [Na/Fe] and [Al/Fe] show no
trends with evolution, as would be expected in the self-pollution
scenario.

\begin{figure}
\begin{center}
\scalebox{.5}[.5]{
\plotone{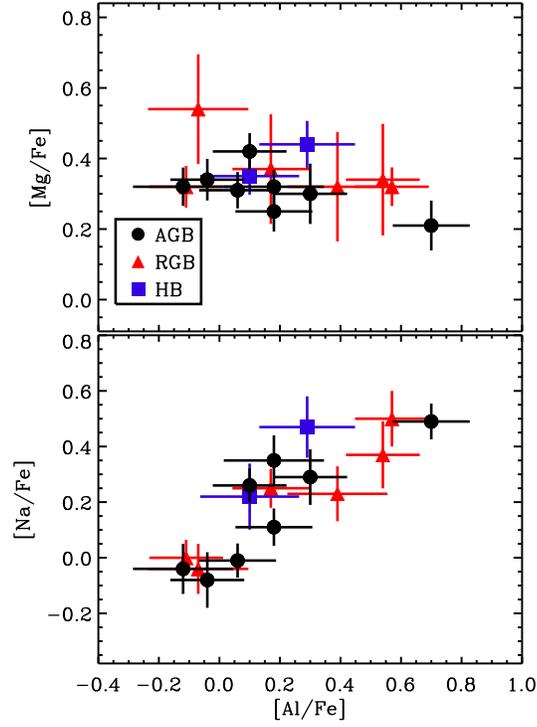}
}
\end{center}
\figcaption{
Relative behavior of both Mg and Na vs. Al abundance for the M5
sample. Two points at [Al/Fe] = $-$0.1 and +0.7 drive the possible appearance of an Mg$-$Al anticorrelation, but 
the Na$-$Al correlation is clear, with little or no offset among stars in different evolutionary states.
For two of our stars, HB434 and HB8, [Al/Fe] could only be derived from using the
Al resonance lines at 3944 and 3962 \AA{}. For these two stars, an
NLTE correction is adopted from \citet{alnlte} of +0.65 dex.
\label{NaAl}}
\end{figure}

An Mg$-$Al anticorrelation, however, arising from the Mg$-$Al cycle is
uncertain. In Figure \ref{NaAl}, the variation in Mg is primarily
driven by two points and within errors is consistent with having no
anticorrelation, agreeing with the findings of
\citet{carretta09b}. Given the discussion in Section 5.1.1, this is
hardly surprising. However, a
strong correlation in Na$-$Al is clearly present in Figure
\ref{NaAl}. Taken with the behavior of the CNO abundances with [Al/Fe]
shown in Figure \ref{NaAlN}, this strongly suggests the presence of
the Mg$-$Al cycle.

\begin{figure}
\begin{center}
\scalebox{.5}[.5]{
\plotone{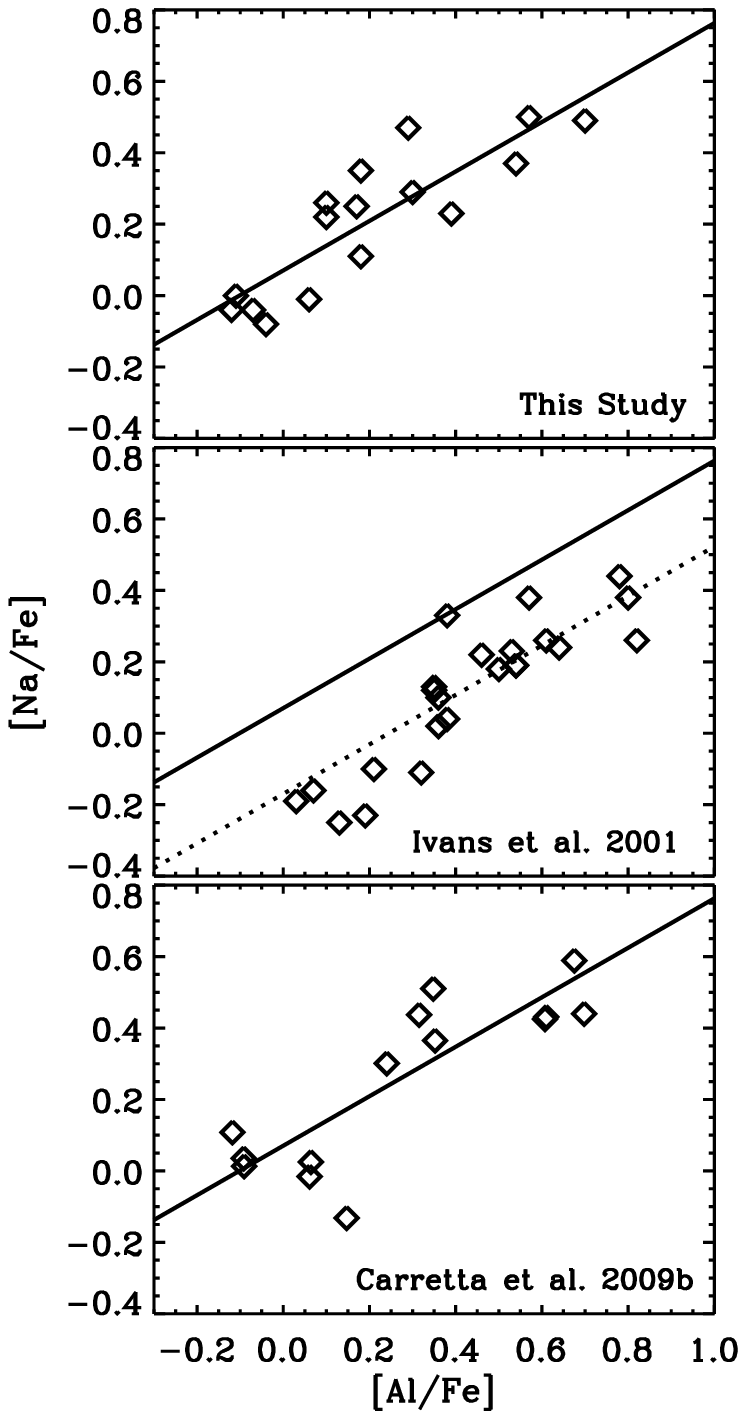}
}
\end{center}
\figcaption{
Na$-$Al correlation from this study is compared to those found by \citet{ivans01} and \citet{carretta09b}.
The solid line in each plot is the best-fit line to our Na$-$Al 
measurements, and matches fairly well the data points of \citet{carretta09b}.
The \citet{ivans01} data lie below our best-fit line; however, this can be almost 
completely accounted for by different adopted \gf{} values for the measured Al
transitions. Assuming the \gf{} values from \citet{ivans01} would reduce our Al
abundances by $\sim 0.24$ dex. Shifting our best-fit line by this
amount gives the dotted line, which agrees very well with the trend seen in the \citet{ivans01} data.
\label{na-alcomp}}
\end{figure}

In Figure \ref{na-alcomp} we show how our Na$-$Al correlation compares
to the previous M5 studies of RGB and AGB stars in \citet{ivans01} and
RGB stars in \citet{carretta09a}. The offset between the correlation found from our
data (plotted as a solid line in \ref{na-alcomp}) and the
\citet{ivans01} values is almost exactly accounted for by different
\gf{} values adopted in our Al abundance determinations.  We updated
the Al \gf{} values used by \citet{ivans01} for the $\lambda 6696$ and
$\lambda 6699$ transitions to current values given by NIST, which are
very similar to the \gf{} values used in \citet{carretta09a}. With the
offset accounted for, the trend we find matches both of the previous
studies very well.

Given the correlations described above, it is likely that the CNO,
Ne$-$Na, and Mg$-$Al cycles operated in a common site that has influenced
the abundances of some fraction of M5 stars. Our data cannot discern
between self-pollution of the observed M5 stars via deep mixing, or a
primordial scenario giving rise to these observational signatures.
However, the studies of unevolved stars in M5 by
\citet{cbs02} and \citet{rc03} find that these relationships exist in
earlier stellar evolutionary branches in M5. With our match to previous work
on the Na-Al correlation, we are consistent with the picture of
multiple generations of stars giving rise to heterogeneous
self-enrichment of the light elements in globular clusters, a scenario
that was introduced in an early form by \citet{cottrell81},
and which has been discussed and expanded upon in a number of papers, such as the
recent work by \citet{carretta10b}.

\subsection{The Neutron-capture Signature}

We are able to measure many neutron-capture transitions in our spectra. This provides an opportunity
to better understand the nucleosynthetic footprint of the
neutron-capture processes that have occurred in the stars of M5. Coupled
with the sample selection of AGB stars, we can also check for any
early onset of $s$-process dredge-up.

\begin{figure}
\begin{center}
\scalebox{.7}[.7]{
\plotone{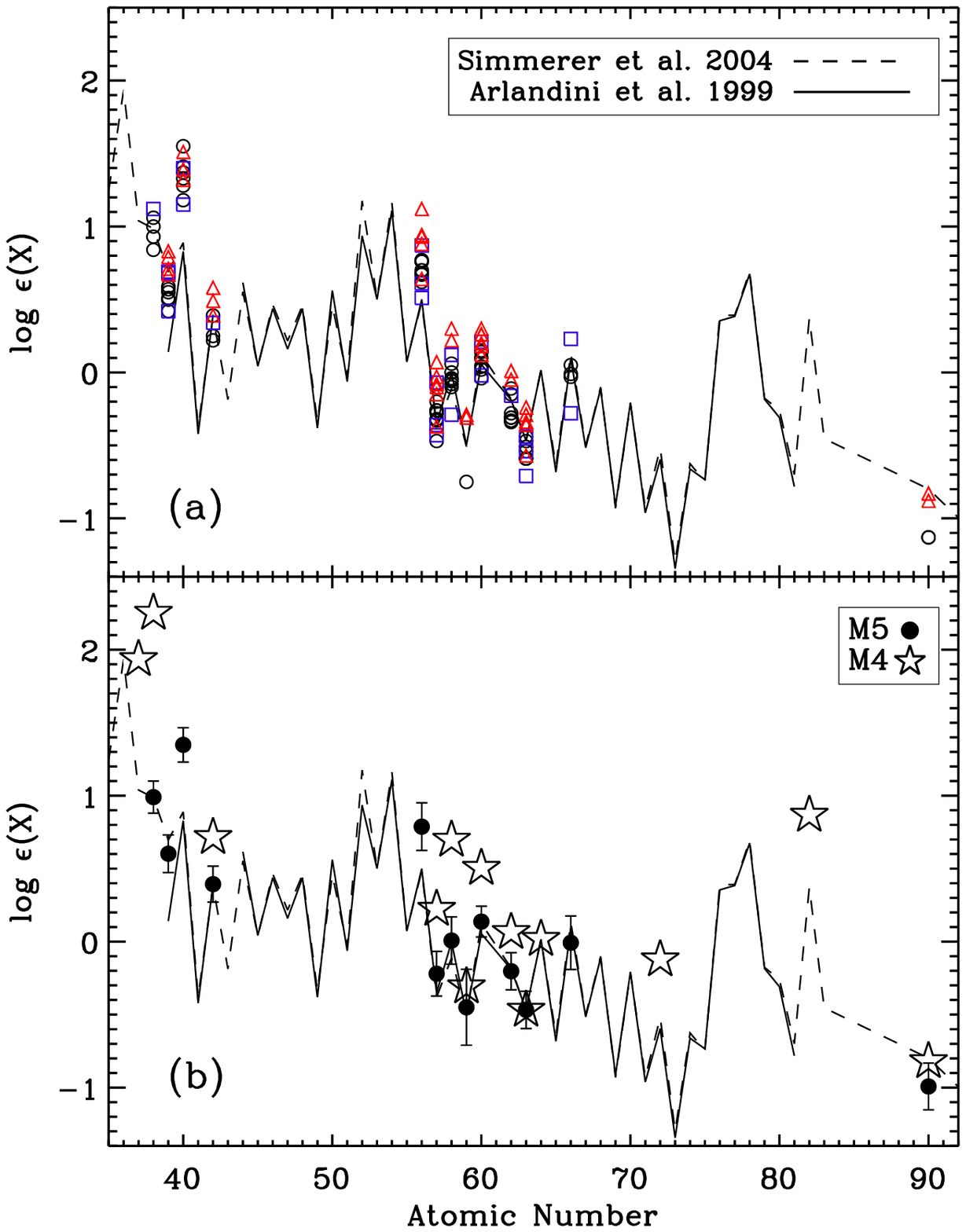} 
}
\end{center}
\figcaption{ (a) Plotted are the measured neutron-capture element
  abundances vs. atomic number for all stars in our M5 sample.  We
  overplot the solar sytem $r$-process abundance patterns from
  \citet{arlandini} and \citet{simmerer04}, scaled to our average
  log$\epsilon$(Eu) value of $-0.47$ ($A=63$). The Zr abundances
  represent both \ion{Zr}{1} and \ion{Zr}{2}. When both ionization
  states of Zr are measured in a star, the average of the two is
  plotted in this figure. The hollow triangles, hollow squares, and
  hollow circles correspond to RGB, HB, and AGB stars (red, blue, and
  black in the electronic edition), respectively. (b) This panel plots
  the average for each abundance measurement (solid circles). The
  error bars represent the standard deviation of the measured
  abundances. Also plotted are the average neutron-capture abundances
  from M4 as measured by \citet{yong08} and \citet{yong08b}, scaled to
  the average log$\epsilon$(Eu) we measure for our M5 sample.
\label{ncapt}}
\end{figure}

In Figure \ref{ncapt}, we plot all of our measured neutron-capture element
abundances and compare them to the scaled solar system $r$-process
pattern. The abundances match the scaled
solar system $r$-process pattern fairly well, and there is also a very
small star-to-star scatter in each abundance. For example, if we take
[Eu/Fe], an element almost entirely synthesized in the $r$-process, we
find an average value of [Eu/Fe]=$0.44 \pm 0.02$ ($\sigma =
0.09$). This average value agrees very well with the larger sample of
\citet{ivans01}, but we find an even smaller scatter than that
study. The lack of star-to-star scatter also falls in line with most
other globular clusters, unlike in the case of M15 \citep{sneden97}.

To attempt to take out the effect of even this small star-to-star
scatter, we compare our neutron-capture abundances relative to Eu in
Figure \ref{ndetails}. For all of these abundance ratios, the values are
virtually identical for all stars and show no difference with
evolutionary state. As shown in Figure \ref{ndetails}, these abundance
ratios match the solar system $r$-process very well, and show no signs
of varying with evolution. There are some
offsets, with the most noticeable exceptions being Zr and Ba. We
conclude that a large majority of all of the neutron-capture
elements in M5 were synthesized in the $r$-process. Figure
\ref{ncapt} shows how this can be contrasted with the globular cluster M4,
which exhibits significant $s$-process contributions to its
neutron-capture elements, even though it is roughly the same
metallicity as M5 \citep{ivans99,yong08}.

\subsubsection{Age}

In Figures \ref{ncapt} and \ref{ndetails}, Th is
low relative to the solar-system $r$-process value, as expected
because it is radioactively unstable. This phenomenon allows us to
date the stars in our sample based on their $\log \epsilon({\rm
  Th/Eu})$ values because Eu is a stable $r$-process element (e.g.,
\citealt{westin,jb01,ivans06,frebel07}). Using an initial $r$-process
production ratio of $\log \epsilon ({\rm Th/Eu}) = -0.28$
\citep{kratz07}, we determine an average age of 12.8 Gyr
($\sigma=2.1$ Gyr) for the four stars with a (Th/Eu) abundance ratio
measurement.

We find Th abundances lower than those reported
by \citet{yong08} for a different M5 sample. Using their $\log \epsilon({\rm
  Th/Eu})$ values, we determine an average cluster age of only 4.2
Gyr. Deriving ages from a (Th/Eu) ratio can be problematic, however,
as this ratio can sometimes be elevated in $r$-process-enhanced
metal-poor stars (e.g., Figure 24 in \citealt{lai08}). An abundance
variation in $\log \epsilon({\rm Th/Eu})$ in M5, though unlikely, may
be one reason for this difference. Another possible cause for the
difference could be in the adopted linelist. However, we use the same
\gf{} value for the 5989 \AA{} \ion{Th}{2} line as \citet{yong08}, and it
is a fairly isolated line. It is possible there are unidentified
features that could affect the synthesis, \citet{aoki07c} point out a
\ion{Nd}{2} line in this region. Including this line, though, did not
affect our results.

Our age determination of 12.8 Gyrs is about 1-3 Gyrs older than other
estimates for the age of M5 based on CMD fitting
\citep{meissner06}. However, given the caveats of using Th as an
age indicator, and using $\sigma=2.1$ Gyr as an estimate of our
error, the Th age result is in reasonable agreement with these
independent measurements.

\subsubsection{Signs of the s-process}

The heightened [Zr/Eu] and [Ba/Eu] ratios seen in our analysis may
indicate an additional nucleosynthetic contribution from the
$s$-process to the stars in M5. For Zr, a partial explanation of our
results may come from comparing the abundances derived from the
ionized species versus those from the neutral species (Table
\ref{abund4}). For the three stars where both are measured, the
neutral state gives an abundance ratio approximately 0.3 dex lower
than the ionized state. This may indicate some systematic abundance
analysis problems with Zr (e.g., as discussed in Section
\ref{results}). Furthermore, interpreting the light neutron-capture
elements including Zr (also Sr, Y, and Mo) is complicated by the
possibility that they can have other nucleosynthetic contributions in
addition to the main $s$-process \citep{sneden08}, including the weak
$s$-process \citep{pignatari08}, charged particle reactions
\citep{qian08}, and a light-element primary process \citep{travaglio}.

\begin{figure}
\begin{center}
\scalebox{1}[1]{
\plotone{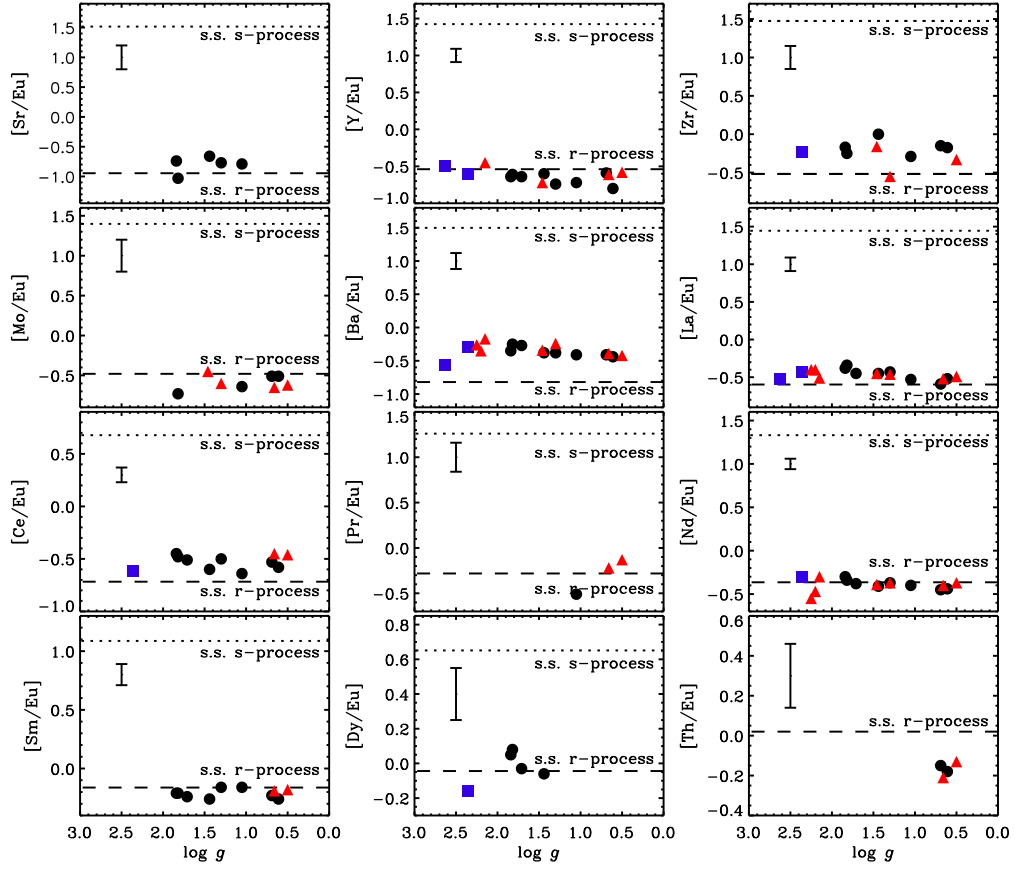}
}
\end{center}
\figcaption{
Heavy element to Eu ratios [X/Eu] measured for the M5 stars in our sample vs. surface gravity.
The $r$-process origin of the neutron-capture elements we measure is
evident when we compare to the solar system $r$-process (dashed lines) and solar system
$s$-process (dotted lines) ratios as derived from
\citet{simmerer04}. The symbols are as in Figure \ref{ca}, i.e., triangles, 
squares, and circles correspond to RGB, HB, and AGB stars,
respectively. A typical error bar is shown in the top left corner of
each plot. The average
values of [X/Eu] and $\sigma$ are given in Table \ref{vsEu}.
\label{ndetails}}
\end{figure}

However, the high measured [Zr/Eu] and [Ba/Eu] abundance ratios
would fit the findings of an $s$-process
contribution in \citet{yong08,yong08b} for their sample of
two M5 RGB stars. In Figure \ref{ndetails}, the nearly constant
[Ba/Eu] and [La/Eu] over all evolutionary states indicate that this
is not from an early onset of $s$-process material being dredged up in
these AGB stars. Closer inspection of Figure \ref{ndetails} does show
an intriguing trend where some of the elements are slightly elevated
relative to the $r$-process-only line, although at lesser degrees than
Zr and Ba.

We explore this in more detail in Table \ref{vsEu}, where we present
the average [X/Eu] abundance ratios of our sample along with the
solar system $r$-process [X/Eu] ratio and percentage contribution to
the solar system abundance from the $r$-process
\citep{simmerer04}. All elements
that can have a significant contribution from the $s$-process ($>70$\%
in the solar system: Sr, Zr, Ba, La, and Ce, with the light-neutron
capture elements Y and Mo being the only exceptions) are elevated in
their [X/Eu] abundance ratio in M5 relative to the solar-system
$r$-process [X/Eu]. On the other hand, the elements that can have a
significant contribution from the $r$-process ($> 30$\% in the solar
system) fall almost exactly on the solar system $r$-process [X/Eu]
value in M5 (Pr, Nd, Sm, and Dy).

\begin{figure}
\begin{center}
\scalebox{1}[1]{
\plotone{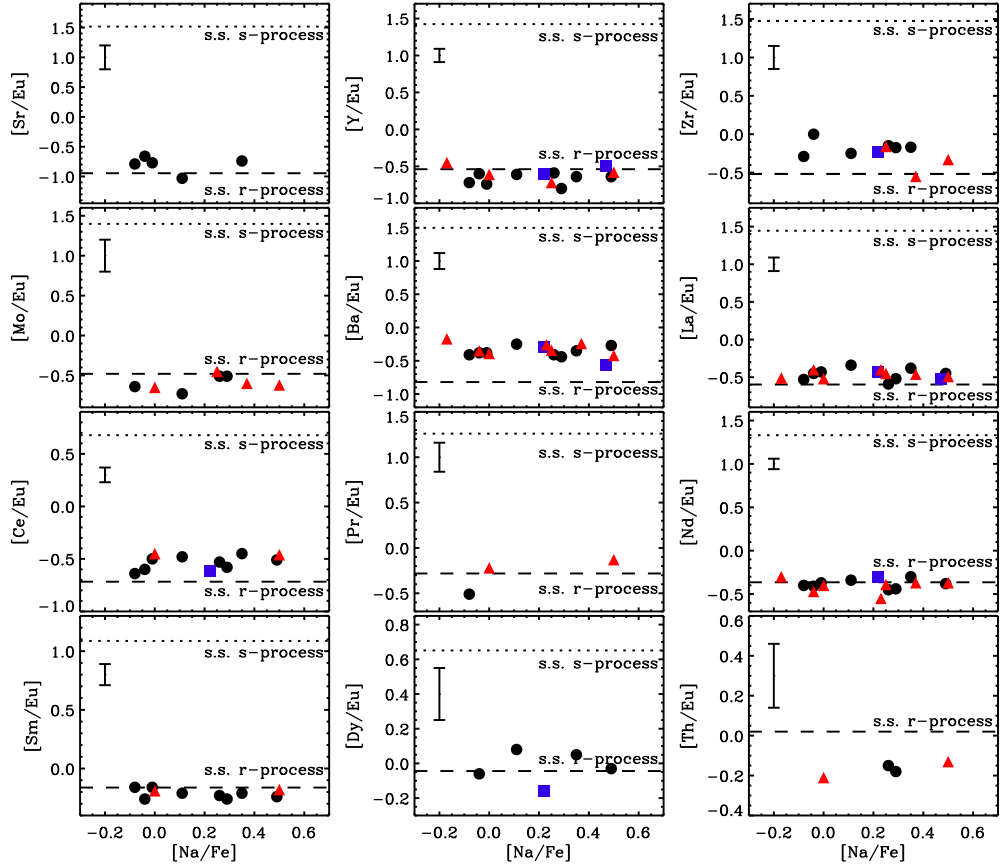}
}
\end{center}
\figcaption{
Heavy element to Eu ratios [X/Eu] measured for the M5 stars in
our sample vs. [Na/Fe]. A typical error bar is shown in the top left corner of
each plot. The lack of correlation with evolution
shown in Figure \ref{ndetails} persists when compared to the
light-element variations such as [Na/Fe]. 
\label{ndetails_na}}
\end{figure}

While clearly not $s$-process-dominated like the globular cluster M4,
our derived abundances are an indication that there was some
$s$-process material produced by an early generation of stars that
contributed to the heavy element content of M5. These stars may have
either formed within M5 and caused cluster self-enrichment, or formed
before M5 and pre-enriched the gas that eventually became incorporated
into its protocluster. While the overall scatter in each
neutron-capture element abundance ratio relative to Eu is very low, we
can test if this low scatter is tied to the light-element
abundance variations. Figure \ref{ndetails_na} reproduces Figure
\ref{ndetails}, but with [X/Eu] versus [Na/Fe]. It is clear from this
plot that there is no correlation of [X/Eu] with [Na/Fe].
Combined with the uniformity of the neutron-capture signature in the
sample, this suggests that the $s$-process enhancements came from the protocluster gas, and not
from cluster self-enrichment. As \citet{yong08b} point out,
intermediate-mass AGB (IM-AGB) stars (in the mass range that
experiences hot bottom burning) will produce very little to no
$s$-process \citep{lattanzio04}, making it possible that they polluted
the protocluster gas.

The sensitivity of the main $s$-process production to AGB-star mass
could be linked to the existence of a smaller amount of $s$-process
material in M5 compared to M4. In the M5 protocluster the mass
function of pre-enriching AGB stars may not have extended to as low a
mass as for the M4 protocluster, with the consequence that
$s$-processing from AGB stars did not contribute as much to M5 as to
M4. Such a circumstance might correlate with M5 being an outer halo
globular cluster, while M4 has Galactic thick-disk-like kinematics. If
M5 formed at an earlier time in Galactic history than M4, then less of
a contribution could have been made to its element enrichment by the
lowest mass IM$-$AGB stars. Alternatively, M5 may have formed in an
$s$-process-poor dwarf ``galaxy''' that was acquired into the Galactic
halo at a very early time in Galactic history, whereas M4 formed from
the more chemically evolved proto-thick-disk. How common
[$s$-process/Fe] variations are, and the level to which they are
present in other $r$-process-dominated clusters, would provide an
important observational constraint to such scenarios.

The important caveat to this interpretation are the potential
systematic offsets that may affect our abundance ratios as discussed
in Section \ref{results}. There is a slight trend with \logg{} in the
[Y/Eu] and [Ba/Eu] abundance. The level of the trends, however, does not affect
the above interpretation. Also, at least when comparing our neutron-capture
abundances relative to Eu, there appears to be no offsets between
stars of different stellar evolutionary stages in our
investigation.  We could, however, assume that the 0.15 dex
offset found in our [V/Fe] measurements also uniformly affects
all of our [X/Eu] determinations. A decrease in the average [La/Eu]
and [Ce/Eu] ratios by this amount would allow these elements to be
attributed to a pure $r$-process. However, it would still be necessary
to explain the excess [Ba/Eu] values, and we would have to assume that
none of the other neutron-capture [X/Eu] values are affected by this
offset in a similar way.

\section{Conclusions}

We have performed detailed abundance analyses for a sample of evolved
stars in the globular cluster M5, covering the RGB, RHB,
and AGB branches of the CMD.
Our conclusions can be broken into three parts.

The first is a cautionary note. It appears that there can be
systematic abundance offsets induced when analyzing stars on the
different evolutionary branches using standard onedimensional LTE abundance
analysis and atmospheres. This can manifest itself not only in
absolute abundances, but more worryingly in abundance ratios such as
[Ca/Fe]. The largest offset found was in [V/Fe], with a
0.15 dex difference between the AGB versus RGB stars. This puts a
limit on how well abundances among the different
evolutionary branches of M5 can be compared.

The second conclusion is that our sample clearly shows the
signatures of star-to-star abundance differences related to the
CNO, Ne$-$Na, and Mg$-$Al nuclear reaction cycles, and that there
are no discernible differences in these element patterns with
stellar evolution phase. This agrees with theoretical predictions that
self-pollution and mixing within present-day globular cluster
stars will not begin until the thermal pulsation phase of
the AGB. Taking this into context with the M5 turn-off stars analyzed
in \citet{rc03}, the (N-,Na-,Al)-rich cluster stars are
present throughout all post-main-sequence phases of evolution. This
is consistent with such stars having acquired their surface element
enhancements from external sources, and not from the
outwards transport of material processed through H-burning reactions
within their interiors.

The third conclusion is that we find the neutron-capture abundances of
M5 to be $r$-process dominated, but with what we interpret as a
small uniform addition of $s$-process material. This neutron-capture
signature is constant through all stars in our sample, depending on
neither evolution nor the light-element variations. This suggests that
low-mass AGB stars contributed heavy elements to the primordial
cluster environment. However, the lack of correlation with the
light-element variations also seems to preclude low-mass AGB stars
from having much of a contribution once star formation had begun.

\acknowledgements

D.K.L acknowledges the support of the National Science Foundation
through the NSF Astronomy and Astrophysics Postdoctoral Fellowship
under award AST-0802292.

G.H.S. acknowledges support from NSF grant AST-0406988. M. B. and
J. A. J. acknowledges support from NSF grant AST-0607770 and
AST-0607482. S.L. is grateful to the DFG cluster of excellence
``Origin and Structure of the Universe'' for partial support. We thank
the anonymous referee whose careful reading and suggestions helped
improve the paper.

{\it Facility:} \facility{Keck:I (HIRES)}

\bibliography{all.bib}

\clearpage

\clearpage

\clearpage

\begin{deluxetable}{llccrclc}
\tablecolumns{8}
\tabletypesize{\small}
\tablewidth{0pc}
\tablecaption{Observation Details \label{obsdetails}}
\tablehead{
       \colhead{Name} &  \colhead{Alt.} & \colhead{$V$} &
       \colhead{$B-V$} & \colhead{Observing} & \colhead{Total} &
       \colhead{S/N at} & \colhead{Evolutionary}\\ 
       \colhead{} & \colhead{Name} & \colhead{(mag)} & \colhead{(mag)}
       & \colhead{Run} & \colhead{Exp. (s)} & \colhead{6035 \AA{}} & \colhead{State}}
\startdata
A9    &         &   12.45   &   1.26  &  2007 Jun &  1200  &  142 &  AGB  \\
A11   &  IV-59  &   12.66   &   1.28  &  2007 Jun &  1980  &  123 &  AGB  \\
A21   &  II-258 &   13.10   &   1.13  &  2007 Jun &  1800  &  131 &  AGB  \\
A25   &         &   13.34   &   0.99  &  2007 Jun &  2100  &  115 &  AGB  \\
A43   &         &   13.64   &   0.96  &  2007 Jun &  2700  &  135 &  AGB  \\
A65   &  I-67   &   13.96   &   0.82  &  2007 Jun &  3600  &  104 &  AGB  \\
A93   &         &   14.23   &   0.79  &  2007 Jun &  5400  &  132 &  AGB  \\
HB13  &         &   14.51   &   0.90  &  2007 Jun &  5760  &  125 &  AGB  \\
HB434 &         &   14.99   &   0.60  &  2007 Jun &  7200  &  113 &  HB   \\
HB8   &         &   15.08   &   0.47  &  2007 Jun &  7200  &  100 &  HB   \\
R9    &         &   12.26   &   1.60  &  2007 Jun &  900   &  115 &  RGB  \\
R21   &         &   12.53   &   1.41  &  2007 Jun &  1800  &  147 &  RGB  \\
R90   &  IV-74  &   13.49   &   1.10  &  2000 Jun &  1800  &  78  &  RGB  \\
R100  &  I-25   &   13.58   &   1.07  &  2000 Jun &  1800  &  105 &  RGB  \\
R394  &  II-16  &   14.99   &   0.86  &  2000 Jun &  7200  &  121 &  RGB  \\
R431  &  IV-24  &   15.09   &   0.85  &  2000 Jun &  5400  &  96  &  RGB  \\
I-65  &         &   15.19   &   0.82  &  2000 Jun &  7200  &  116 &  RGB  
\enddata
\end{deluxetable}

\begin{deluxetable}{lrrrrrrrr}
\tablecolumns{9}
\tabletypesize{\small}
\tablewidth{0pc}
\tablecaption{Atomic Parameters and Equivalent Widths \label{EW}}
\tablehead{
\colhead{Wavelength} & \colhead{Element} & \colhead{  log $gf$} &
\colhead{     EP} & \colhead{ Ref.} & \colhead{   A9} & \colhead{  A11} & \colhead{  A21} & \colhead{  A25} 
}
\startdata
  6300.30 &      8.0 &    $-$9.78 &     0.00 &  1 &   \nodata &     \nodata &       51.8  &     \nodata      \\ 
  6363.78 &      8.0 &   $-$10.30 &     0.02 &  1 &    18.8  &       20.6  &       18.9  &     \nodata      \\ 
  7771.94 &      8.0 &     0.37 &     9.15 &  1 &  \nodata &       14.2  &       19.5  &       17.9       \\ 
  7774.17 &      8.0 &     0.22 &     9.15 &  1 &     8.5  &       12.5  &       14.9  &       18.0       \\ 
  7775.39 &      8.0 &     0.00 &     9.15 &  1 &  \nodata &        6.0  &        8.9  &       11.6       \\
  5682.65 &     11.0 &    $-$0.70 &     2.10 &  2 &    78.40 &       68.40 &       33.60 &       32.00    
\enddata
\tablecomments{Table \ref{EW} is published in its entirety in
the electronic edition. A portion is
shown here for guidance regarding its form and content.}
\tablerefs{
(1) \citealt{rc03}; (2) \citealt{ivans06}; (3) \citealt{nist}; (4) \citealt{mgtriplet}; (5) \citealt{sneden03}; (6) \citealt{sobeck07}; (7) \citealt{crIIlines}; (8) \citealt{felines}; (9) \citealt{kurucz95};
(10) \citealt{ljung}; (11) \citealt{yong08}
}
\end{deluxetable}

\begin{deluxetable}{lccccc}
\tablecolumns{6}
\tabletypesize{\small}
\tablewidth{0pc}
\tablecaption{Stellar Parameters and Radial Velocities\label{parameters}}
\tablehead{
       \colhead{Name} &  \colhead{\teff{}} & \colhead{\logg{}} &
       \colhead{$v_t$} & \colhead{[A/H]} & \colhead{Radial}\\ 
       \colhead{} & \colhead{(K)} & \colhead{} & \colhead{(km s$^{-1}$)}
       & \colhead{} & \colhead{Velocity (km s$^{-1}$)}
}
\startdata
A9    &   4236   &  0.61  &  2.28  & $-1.50$ &  50.5  \\
A11   &   4209   &  0.69  &  2.23  & $-1.50$ &  60.0  \\
A21   &   4381   &  1.05  &  1.82  & $-1.50$ &  47.4  \\
A25   &   4584   &  1.30  &  1.88  & $-1.50$ &  52.5  \\
A43   &   4630   &  1.44  &  1.85  & $-1.50$ &  53.3  \\
A65   &   4893   &  1.71  &  1.95  & $-1.50$ &  54.9  \\
A93   &   4961   &  1.84  &  1.87  & $-1.50$ &  48.7  \\
HB13  &   4740   &  1.82  &  1.35  & $-1.50$ &  48.9  \\
HB434 &   5400   &  2.36  &  2.06  & $-1.50$ &  58.8  \\
HB8   &   6100   &  2.63  &  3.38  & $-1.60$ &  51.5  \\
R9    &   4000   &  0.50  &  1.86  & $-1.35$ &  54.7  \\
R21   &   4100   &  0.66  &  1.81  & $-1.35$ &  52.2  \\
R90   &   4475   &  1.30  &  1.55  & $-1.40$ &  57.2  \\
R100  &   4525   &  1.46  &  1.55  & $-1.40$ &  58.1  \\
R394  &   4818   &  2.15  &  1.27  & $-1.40$ &  51.7  \\
R431  &   4845   &  2.20  &  1.24  & $-1.35$ &  47.9  \\
I-65  &   4860   &  2.25  &  1.20  & $-1.30$ &  58.5
\enddata
\end{deluxetable}

\begin{center}

\begin{deluxetable}{lrrrrrrrrrrrrrrrr}
\tablecolumns{15}
\tabletypesize{\small}
\tablecaption{Abundance Ratios [Fe/H] Through [O/Fe] \label{abund1}}
\tablewidth{0pc}
\tablehead{ \colhead{Star ID} & \colhead{   [Fe/H]} &  \colhead{$N$} & \colhead{ [FeII/H]} &  \colhead{$N$} & \colhead{ log$\epsilon$(Li)} &  \colhead{$N$} & \colhead{$^{12}$C/$^{13}$C} &  \colhead{$N$} & \colhead{   [C/Fe]} &  \colhead{$N$} & \colhead{   [N/Fe]} &  \colhead{$N$} & \colhead{   [O/Fe]} &  \colhead{$N$}  }
\startdata
  A11 & $ -1.56 \pm 0.11 $ & 200 & $ -1.47 \pm 0.09 $ &  12 &    \nodata         &   0 & $  5.00 \pm 2.00 $ &   2 & $  0.14 \pm 0.33 $ &   1 & $  1.16 \pm 0.42 $ &   1 & $  0.49 \pm 0.08 $ &   4\\
  A21 & $ -1.53 \pm 0.11 $ & 208 & $ -1.45 \pm 0.08 $ &  19 &    \nodata         &   0 & $  4.50 \pm 2.00 $ &   2 & $ -0.17 \pm 0.33 $ &   1 & $  0.43 \pm 0.36 $ &   1 & $  0.47 \pm 0.03 $ &   5\\
  A25 & $ -1.50 \pm 0.11 $ & 203 & $ -1.46 \pm 0.12 $ &  17 &    \nodata         &   0 & $  5.00 \pm 2.00 $ &   2 & $ -0.27 \pm 0.17 $ &   1 & $  0.35 \pm 0.23 $ &   1 & $  0.31 \pm 0.11 $ &   3\\
  A43 & $ -1.56 \pm 0.11 $ & 228 & $ -1.43 \pm 0.11 $ &  26 &    \nodata         &   0 & $  6.00 \pm 2.00 $ &   2 & $ -0.04 \pm 0.17 $ &   2 & $  0.31 \pm 0.23 $ &   1 & $  0.56 \pm 0.12 $ &   4\\
  A65 & $ -1.66 \pm 0.11 $ & 144 & $ -1.43 \pm 0.11 $ &  19 &    \nodata         &   0 &        \nodata     &   0 & $ -0.79 \pm 0.17 $ &   2 & $  1.36 \pm 0.23 $ &   1 & $  0.02 \pm 0.19 $ &   2\\
   A9 & $ -1.49 \pm 0.11 $ & 200 & $ -1.49 \pm 0.09 $ &  12 &    \nodata         &   0 & $  6.00 \pm 2.00 $ &   1 & $ -0.51 \pm 0.30 $ &   2 & $  0.99 \pm 0.36 $ &   1 & $  0.28 \pm 0.11 $ &   2\\
  A93 & $ -1.59 \pm 0.11 $ & 176 & $ -1.48 \pm 0.11 $ &  25 &    \nodata         &   0 & $  5.00 \pm 2.00 $ &   1 & $ -0.41 \pm 0.17 $ &   2 & $  0.89 \pm 0.23 $ &   1 & $  0.30 \pm 0.10 $ &   4\\
 HB13 & $ -1.37 \pm 0.11 $ & 222 & $ -1.47 \pm 0.11 $ &  24 &    \nodata         &   0 &        \nodata     &   0 & $ -0.25 \pm 0.22 $ &   1 & $  0.57 \pm 0.23 $ &   1 & $  0.16 \pm 0.11 $ &   3\\
HB434 & $ -1.82 \pm 0.10 $ & 150 & $ -1.53 \pm 0.08 $ &  23 &    \nodata         &   0 &        \nodata     &   0 & $ -0.18 \pm 0.31 $ &   1 & $  < 0.92        $ &   1 & $  0.72 \pm 0.16 $ &   3\\
  HB8 & $ -1.77 \pm 0.10 $ &  83 & $ -1.60 \pm 0.08 $ &  13 &    \nodata         &   0 &        \nodata     &   0 & $   < 0.50       $ &   1 & $  < 1.70        $ &   1 & $  0.58 \pm 0.12 $ &   3\\
  R21 & $ -1.30 \pm 0.11 $ & 215 & $ -1.32 \pm 0.09 $ &  14 &    \nodata         &   0 & $  7.00 \pm 2.00 $ &   1 & $ -0.17 \pm 0.33 $ &   1 & $ -0.05 \pm 0.36 $ &   1 & $  0.29 \pm 0.06 $ &   5\\
   R9 & $ -1.38 \pm 0.11 $ & 192 & $ -1.34 \pm 0.09 $ &  11 &    \nodata         &   0 & $  6.00 \pm 2.00 $ &   1 & $ -0.34 \pm 0.33 $ &   1 & $  0.33 \pm 0.42 $ &   1 & $ -0.05 \pm 0.15 $ &   1\\
 R100 & $ -1.40 \pm 0.11 $ & 170 & $ -1.36 \pm 0.12 $ &   8 &    \nodata         &   0 &        \nodata     &   0 &      \nodata       &   0 &      \nodata       &   0 & $  0.25 \pm 0.16 $ &   1\\
 R394 & $ -1.51 \pm 0.11 $ & 147 & $ -1.35 \pm 0.13 $ &   8 & $  0.76 \pm 0.21 $ &   1 &        \nodata     &   0 &      \nodata       &   0 &      \nodata       &   0 & $  0.14 \pm 0.16 $ &   1\\
 R431 & $ -1.46 \pm 0.11 $ & 144 & $ -1.30 \pm 0.12 $ &   7 & $  0.96 \pm 0.21 $ &   1 &        \nodata     &   0 &      \nodata       &   0 &      \nodata       &   0 & $  0.10 \pm 0.16 $ &   1\\
  R90 & $ -1.38 \pm 0.11 $ & 164 & $ -1.45 \pm 0.09 $ &   7 &    \nodata         &   0 &        \nodata     &   0 &      \nodata       &   0 &      \nodata       &   0 &      \nodata       &   0\\
 I-65 & $ -1.40 \pm 0.11 $ & 136 & $ -1.22 \pm 0.12 $ &   8 & $  0.73 \pm 0.21 $ &   1 &        \nodata     &   0 &      \nodata       &   0 &      \nodata       &   0 &      \nodata       &   0
\enddata
\end{deluxetable}

\begin{deluxetable}{lrrrrrrrrrrrrrrrr}
\tablecolumns{15}
\tabletypesize{\small}
\tablecaption{Abundance Ratios [Na/Fe] Through [\ion{Ti}{1}/Fe] \label{abund2}}
\tablewidth{0pc}
\tablehead{ \colhead{Star Id} & \colhead{  [Na/Fe]} &  \colhead{$N$} & \colhead{  [Mg/Fe]} &  \colhead{$N$} & \colhead{  [Al/Fe]} &  \colhead{$N$} & \colhead{  [Si/Fe]} &  \colhead{$N$} & \colhead{  [Ca/Fe]} &  \colhead{$N$} & \colhead{[ScII/Fe]} &  \colhead{$N$} & \colhead{ [TiI/Fe]} &  \colhead{$N$}  }
\startdata
  A11 & $  0.26 \pm 0.06 $ &   5 & $  0.42 \pm 0.05 $ &   5 & $  0.10 \pm 0.12 $ &   2 & $  0.48 \pm 0.11 $ &  13 & $  0.15 \pm 0.05 $ &  13 & $  0.04 \pm 0.07 $ &   6 & $  0.09 \pm 0.07 $ &  46\\
  A21 & $ -0.08 \pm 0.10 $ &   3 & $  0.34 \pm 0.06 $ &   5 & $ -0.04 \pm 0.12 $ &   2 & $  0.40 \pm 0.11 $ &  15 & $  0.23 \pm 0.05 $ &  14 & $  0.08 \pm 0.07 $ &   7 & $  0.08 \pm 0.07 $ &  48\\
  A25 & $ -0.01 \pm 0.06 $ &   5 & $  0.31 \pm 0.05 $ &   6 & $  0.06 \pm 0.13 $ &   2 & $  0.39 \pm 0.08 $ &  15 & $  0.23 \pm 0.05 $ &  17 & $  0.13 \pm 0.06 $ &   9 & $  0.17 \pm 0.06 $ &  38\\
  A43 & $ -0.04 \pm 0.09 $ &   4 & $  0.32 \pm 0.05 $ &   6 & $ -0.12 \pm 0.17 $ &   1 & $  0.43 \pm 0.09 $ &  16 & $  0.25 \pm 0.04 $ &  16 & $  0.08 \pm 0.06 $ &  10 & $  0.17 \pm 0.06 $ &  43\\
  A65 & $  0.49 \pm 0.06 $ &   5 & $  0.21 \pm 0.07 $ &   5 & $  0.70 \pm 0.13 $ &   2 & $  0.51 \pm 0.09 $ &  13 & $  0.33 \pm 0.04 $ &  16 & $  0.10 \pm 0.05 $ &   9 & $  0.13 \pm 0.07 $ &  19\\
   A9 & $  0.29 \pm 0.10 $ &   3 & $  0.30 \pm 0.09 $ &   6 & $  0.30 \pm 0.12 $ &   2 & $  0.39 \pm 0.12 $ &  14 & $  0.18 \pm 0.06 $ &  10 & $  0.13 \pm 0.08 $ &   7 & $  0.17 \pm 0.07 $ &  41\\
  A93 & $  0.35 \pm 0.09 $ &   4 & $  0.32 \pm 0.05 $ &   7 & $  0.18 \pm 0.17 $ &   1 & $  0.44 \pm 0.09 $ &  14 & $  0.26 \pm 0.04 $ &  18 & $  0.15 \pm 0.05 $ &  10 & $  0.18 \pm 0.06 $ &  29\\
 HB13 & $  0.11 \pm 0.07 $ &   6 & $  0.25 \pm 0.06 $ &   6 & $  0.18 \pm 0.13 $ &   2 & $  0.30 \pm 0.08 $ &  16 & $  0.31 \pm 0.04 $ &  19 & $  0.28 \pm 0.05 $ &  10 & $  0.21 \pm 0.06 $ &  50\\
HB434 & $  0.22 \pm 0.12 $ &   4 & $  0.35 \pm 0.05 $ &   7 & $  0.10 \pm 0.16 $ &   2 & $  0.61 \pm 0.11 $ &   3 & $  0.36 \pm 0.05 $ &  18 & $  0.07 \pm 0.07 $ &   9 & $  0.25 \pm 0.03 $ &  20\\
  HB8 & $  0.47 \pm 0.11 $ &   2 & $  0.44 \pm 0.07 $ &   6 & $  0.29 \pm 0.16 $ &   1 & $  0.32 \pm 0.12 $ &   2 & $  0.32 \pm 0.04 $ &  12 & $  0.23 \pm 0.07 $ &   5 & $  0.31 \pm 0.09 $ &   3\\
  R21 & $  0.00 \pm 0.06 $ &   5 & $  0.32 \pm 0.06 $ &   5 & $ -0.11 \pm 0.12 $ &   2 & $  0.29 \pm 0.11 $ &  15 & $  0.25 \pm 0.05 $ &  14 & $  0.21 \pm 0.06 $ &   6 & $  0.26 \pm 0.07 $ &  48\\
   R9 & $  0.50 \pm 0.10 $ &   3 & $  0.32 \pm 0.05 $ &   5 & $  0.57 \pm 0.12 $ &   2 & $  0.42 \pm 0.12 $ &  14 & $  0.24 \pm 0.06 $ &  13 & $  0.16 \pm 0.06 $ &   5 & $  0.19 \pm 0.07 $ &  42\\
 R100 & $  0.25 \pm 0.07 $ &   5 & $  0.37 \pm 0.16 $ &   1 & $  0.17 \pm 0.13 $ &   2 & $  0.38 \pm 0.09 $ &  15 & $  0.35 \pm 0.04 $ &  17 & $  0.23 \pm 0.05 $ &   7 & $  0.23 \pm 0.07 $ &  25\\
 R394 & $ -0.17 \pm 0.10 $ &   3 & $  0.35 \pm 0.16 $ &   1 &    \nodata         &   0 & $  0.34 \pm 0.09 $ &  18 & $  0.35 \pm 0.04 $ &  18 & $  0.12 \pm 0.05 $ &   7 & $  0.24 \pm 0.07 $ &  12\\
 R431 & $ -0.04 \pm 0.09 $ &   4 & $  0.54 \pm 0.16 $ &   1 & $ -0.07 \pm 0.17 $ &   1 & $  0.39 \pm 0.09 $ &  16 & $  0.37 \pm 0.04 $ &  18 & $  0.10 \pm 0.05 $ &   7 & $  0.23 \pm 0.07 $ &  15\\
  R90 & $  0.37 \pm 0.12 $ &   3 & $  0.34 \pm 0.16 $ &   1 & $  0.54 \pm 0.12 $ &   2 & $  0.39 \pm 0.11 $ &  17 & $  0.35 \pm 0.05 $ &  17 & $  0.32 \pm 0.06 $ &   7 & $  0.18 \pm 0.07 $ &  24\\
 I-65 & $  0.23 \pm 0.10 $ &   4 & $  0.32 \pm 0.16 $ &   1 & $  0.39 \pm 0.17 $ &   1 & $  0.41 \pm 0.09 $ &  14 & $  0.36 \pm 0.04 $ &  18 & $  0.15 \pm 0.05 $ &   8 & $  0.22 \pm 0.06 $ &  13
\enddata
\end{deluxetable}

\begin{deluxetable}{lrrrrrrrrrrrrrrrr}
\tablecolumns{15}
\tabletypesize{\small}
\tablecaption{Abundance Ratios [TiII/Fe] Through [Co/Fe] \label{abund3}}
\tablewidth{0pc}
\tablehead{ \colhead{Star Id} & \colhead{[TiII/Fe]} &  \colhead{$N$} & \colhead{  [VI/Fe]} &  \colhead{$N$} & \colhead{ [VII/Fe]} &  \colhead{$N$} & \colhead{ [CrI/Fe]} &  \colhead{$N$} & \colhead{[CrII/Fe]} &  \colhead{$N$} & \colhead{  [Mn/Fe]} &  \colhead{$N$} & \colhead{  [Co/Fe]} &  \colhead{$N$} }
\startdata
  A11 & $  0.18 \pm 0.07 $ &   6 & $ -0.22 \pm 0.11 $ &  12 &    \nodata         &   0 & $ -0.14 \pm 0.06 $ &  24 & $  0.03 \pm 0.11 $ &   2 & $ -0.36 \pm 0.07 $ &   8 & $  0.00 \pm 0.07 $ &   7\\
  A21 & $  0.26 \pm 0.08 $ &  12 & $ -0.23 \pm 0.11 $ &  12 &    \nodata         &   0 & $ -0.10 \pm 0.05 $ &  17 & $  0.05 \pm 0.10 $ &   3 & $ -0.38 \pm 0.07 $ &   8 & $ -0.11 \pm 0.09 $ &   4\\
  A25 & $  0.26 \pm 0.06 $ &  14 & $ -0.17 \pm 0.09 $ &   9 &    \nodata         &   0 & $ -0.13 \pm 0.05 $ &  18 & $  0.01 \pm 0.07 $ &   5 & $ -0.34 \pm 0.06 $ &   7 & $  0.12 \pm 0.14 $ &   3\\
  A43 & $  0.30 \pm 0.06 $ &  21 & $ -0.23 \pm 0.09 $ &  15 &    \nodata         &   0 & $ -0.12 \pm 0.06 $ &  20 & $ -0.01 \pm 0.05 $ &   8 & $ -0.41 \pm 0.05 $ &   9 & $ -0.02 \pm 0.11 $ &   5\\
  A65 & $  0.33 \pm 0.07 $ &  18 & $ -0.29 \pm 0.12 $ &   3 &    \nodata         &   0 & $ -0.25 \pm 0.03 $ &  12 & $  0.01 \pm 0.08 $ &   4 & $ -0.43 \pm 0.06 $ &   5 & $ -0.02 \pm 0.16 $ &   1\\
   A9 & $  0.24 \pm 0.09 $ &   7 & $ -0.23 \pm 0.14 $ &  10 &    \nodata         &   0 & $ -0.11 \pm 0.05 $ &  18 & $  0.02 \pm 0.16 $ &   1 & $ -0.33 \pm 0.08 $ &   8 & $  0.07 \pm 0.08 $ &   6\\
  A93 & $  0.35 \pm 0.06 $ &  25 & $ -0.26 \pm 0.08 $ &   7 &    \nodata         &   0 & $ -0.17 \pm 0.05 $ &  19 & $  0.02 \pm 0.05 $ &   9 & $ -0.41 \pm 0.06 $ &   7 & $ -0.09 \pm 0.16 $ &   1\\
 HB13 & $  0.39 \pm 0.06 $ &  20 & $ -0.12 \pm 0.09 $ &  10 &    \nodata         &   0 & $  0.02 \pm 0.04 $ &  26 & $  0.14 \pm 0.06 $ &   6 & $ -0.34 \pm 0.05 $ &   8 & $ -0.17 \pm 0.16 $ &   2\\
HB434 & $  0.20 \pm 0.05 $ &  32 & $ -0.17 \pm 0.11 $ &   2 & $ -0.05 \pm 0.17 $ &   1 & $ -0.17 \pm 0.05 $ &  15 & $ -0.01 \pm 0.05 $ &   8 & $ -0.30 \pm 0.06 $ &   8 & $ -0.08 \pm 0.11 $ &   2\\
  HB8 & $  0.39 \pm 0.05 $ &  24 &    \nodata         &   0 & $  0.10 \pm 0.17 $ &   1 & $ -0.04 \pm 0.07 $ &   7 & $  0.05 \pm 0.08 $ &   6 & $ -0.45 \pm 0.11 $ &   2 & $  0.03 \pm 0.16 $ &   1\\
  R21 & $  0.40 \pm 0.08 $ &   8 & $ -0.10 \pm 0.11 $ &  11 &    \nodata         &   0 & $  0.00 \pm 0.05 $ &  33 & $  0.17 \pm 0.16 $ &   1 & $ -0.35 \pm 0.09 $ &   4 & $  0.03 \pm 0.08 $ &   6\\
   R9 & $  0.29 \pm 0.10 $ &   6 & $ -0.11 \pm 0.12 $ &  10 &    \nodata         &   0 & $  0.00 \pm 0.05 $ &  23 & $  0.11 \pm 0.11 $ &   2 & $ -0.51 \pm 0.11 $ &   2 & $  0.05 \pm 0.08 $ &   6\\
 R100 & $  0.32 \pm 0.09 $ &   4 & $  0.02 \pm 0.09 $ &   8 &    \nodata         &   0 & $  0.00 \pm 0.06 $ &  10 & $  0.03 \pm 0.15 $ &   1 & $ -0.38 \pm 0.04 $ &   5 & $  0.01 \pm 0.11 $ &   2\\
 R394 & $  0.21 \pm 0.10 $ &   3 & $ -0.11 \pm 0.08 $ &   8 &    \nodata         &   0 & $ -0.05 \pm 0.06 $ &   7 & $ -0.06 \pm 0.15 $ &   1 & $ -0.43 \pm 0.08 $ &   4 & $ -0.11 \pm 0.16 $ &   1\\
 R431 & $  0.27 \pm 0.10 $ &   3 & $ -0.11 \pm 0.11 $ &   4 &    \nodata         &   0 & $ -0.02 \pm 0.05 $ &   8 & $ -0.14 \pm 0.15 $ &   1 & $ -0.41 \pm 0.08 $ &   4 & $  0.15 \pm 0.11 $ &   2\\
  R90 & $  0.29 \pm 0.10 $ &   3 & $ -0.01 \pm 0.11 $ &  11 &    \nodata         &   0 & $  0.01 \pm 0.06 $ &   9 & $  0.08 \pm 0.16 $ &   1 & $ -0.37 \pm 0.09 $ &   4 & $  0.01 \pm 0.08 $ &   5\\
 I-65 & $  0.32 \pm 0.10 $ &   3 & $ -0.10 \pm 0.09 $ &   6 &    \nodata         &   0 & $  0.03 \pm 0.07 $ &   7 & $ -0.08 \pm 0.15 $ &   1 & $ -0.49 \pm 0.04 $ &   5 & $  0.17 \pm 0.16 $ &   1
\enddata
\end{deluxetable}

\begin{deluxetable}{lrrrrrrrrrrrrrrrr}
\tablecolumns{15}
\tabletypesize{\small}
\tablecaption{Abundance Ratios [Ni/Fe] Through [ZrII/Fe] \label{abund4}}
\tablewidth{0pc}
\tablehead{ \colhead{Star Id} & \colhead{  [Ni/Fe]} &  \colhead{$N$} & \colhead{  [Zn/Fe]} &  \colhead{$N$} & \colhead{  [Cu/Fe]} &  \colhead{$N$} & \colhead{ [SrI/Fe]} &  \colhead{$N$} & \colhead{ [YII/Fe]} &  \colhead{$N$} & \colhead{ [ZrI/Fe]} &  \colhead{$N$} & \colhead{[ZrII/Fe]} &  \colhead{$N$} }
\startdata
  A11 & $ -0.06 \pm 0.03 $ &  40 & $  0.36 \pm 0.27 $ &   2 & $ -0.84 \pm 0.11 $ &   2 &    \nodata         &   0 & $ -0.22 \pm 0.07 $ &   5 & $  0.04 \pm 0.16 $ &   4 & $  0.31 \pm 0.18 $ &   2\\
  A21 & $ -0.08 \pm 0.03 $ &  36 & $  0.16 \pm 0.19 $ &   2 & $ -0.87 \pm 0.11 $ &   2 & $ -0.37 \pm 0.17 $ &   1 & $ -0.22 \pm 0.06 $ &   8 & $ -0.04 \pm 0.16 $ &   3 & $  0.38 \pm 0.11 $ &   3\\
  A25 & $ -0.07 \pm 0.05 $ &  32 & $  0.05 \pm 0.15 $ &   2 & $ -0.87 \pm 0.13 $ &   2 & $ -0.34 \pm 0.17 $ &   1 & $ -0.27 \pm 0.09 $ &   4 &    \nodata         &   0 & $  0.23 \pm 0.11 $ &   2\\
  A43 & $ -0.08 \pm 0.05 $ &  35 & $  0.17 \pm 0.15 $ &   2 & $ -0.89 \pm 0.13 $ &   2 & $ -0.41 \pm 0.17 $ &   1 & $ -0.22 \pm 0.05 $ &   7 &    \nodata         &   0 & $  0.38 \pm 0.10 $ &   4\\
  A65 & $ -0.16 \pm 0.07 $ &  12 & $  0.25 \pm 0.15 $ &   2 & $ -0.94 \pm 0.17 $ &   1 &    \nodata         &   0 & $ -0.31 \pm 0.06 $ &   6 &    \nodata         &   0 &    \nodata         &   0\\
   A9 & $ -0.07 \pm 0.03 $ &  37 & $  0.07 \pm 0.22 $ &   1 & $ -0.87 \pm 0.11 $ &   2 &    \nodata         &   0 & $ -0.33 \pm 0.10 $ &   4 & $  0.11 \pm 0.16 $ &   4 & $  0.48 \pm 0.16 $ &   1\\
  A93 & $ -0.07 \pm 0.05 $ &  21 & $  0.11 \pm 0.15 $ &   2 & $ -0.92 \pm 0.13 $ &   2 & $ -0.47 \pm 0.17 $ &   1 & $ -0.26 \pm 0.05 $ &   5 &    \nodata         &   0 & $  0.21 \pm 0.05 $ &   5\\
 HB13 & $ -0.10 \pm 0.05 $ &  28 & $  0.15 \pm 0.14 $ &   3 & $ -0.83 \pm 0.13 $ &   2 & $ -0.41 \pm 0.17 $ &   1 & $ -0.09 \pm 0.05 $ &   6 &    \nodata         &   0 & $  0.27 \pm 0.11 $ &   2\\
HB434 & $ -0.08 \pm 0.08 $ &   4 & $  0.18 \pm 0.12 $ &   2 & $ -0.88 \pm 0.16 $ &   1 &    \nodata         &   0 & $ -0.29 \pm 0.07 $ &   5 &    \nodata         &   0 & $  0.08 \pm 0.10 $ &   4\\
  HB8 & $ -0.22 \pm 0.09 $ &   3 &    \nodata         &   0 & $  < -0.33       $ &   1 &    \nodata         &   0 & $  0.05 \pm 0.17 $ &   1 &    \nodata         &   0 &    \nodata         &   0\\
  R21 & $ -0.09 \pm 0.03 $ &  44 & $  0.06 \pm 0.19 $ &   2 & $ -0.93 \pm 0.15 $ &   1 &    \nodata         &   0 & $ -0.09 \pm 0.13 $ &   4 & $  0.09 \pm 0.16 $ &   4 &    \nodata         &   0\\
   R9 & $ -0.06 \pm 0.03 $ &  39 &    \nodata         &   0 & $ -0.92 \pm 0.15 $ &   1 &    \nodata         &   0 & $ -0.11 \pm 0.12 $ &   3 & $  0.10 \pm 0.16 $ &   3 &    \nodata         &   0\\
 R100 & $ -0.03 \pm 0.05 $ &  27 &    \nodata         &   0 & $ -0.94 \pm 0.17 $ &   1 &    \nodata         &   0 & $ -0.21 \pm 0.16 $ &   1 & $  0.31 \pm 0.16 $ &   2 &    \nodata         &   0\\
 R394 & $ -0.05 \pm 0.05 $ &  25 &    \nodata         &   0 & $ -1.04 \pm 0.17 $ &   1 &    \nodata         &   0 & $ -0.18 \pm 0.16 $ &   1 &    \nodata         &   0 &    \nodata         &   0\\
 R431 & $ -0.04 \pm 0.05 $ &  23 &    \nodata         &   0 & $ -0.94 \pm 0.17 $ &   1 &    \nodata         &   0 &    \nodata         &   0 &    \nodata         &   0 &    \nodata         &   0\\
  R90 & $ -0.03 \pm 0.04 $ &  25 &    \nodata         &   0 & $ -0.99 \pm 0.15 $ &   1 &    \nodata         &   0 &    \nodata         &   0 & $  0.07 \pm 0.21 $ &   1 &    \nodata         &   0\\
 I-65 & $  0.00 \pm 0.05 $ &  21 &    \nodata         &   0 & $ -1.05 \pm 0.17 $ &   1 &    \nodata         &   0 &    \nodata         &   0 &    \nodata         &   0 &    \nodata         &   0
\enddata
\end{deluxetable}
\end{center}

\begin{deluxetable}{lrrrrrrrrrrrr}
\tablecolumns{11}
\tabletypesize{\small}
\tablecaption{Abundance Ratios [Mo/Fe] Through [PrII/Fe] \label{abund5}}
\tablewidth{0pc}
\tablehead{ \colhead{Star ID} & \colhead{  [Mo/Fe]} &  \colhead{$N$} & \colhead{[BaII/Fe]} &  \colhead{$N$} & \colhead{[LaII/Fe]} &  \colhead{$N$} & \colhead{[CeII/Fe]} &  \colhead{$N$} & \colhead{[PrII/Fe]} &  \colhead{$N$}  }
\startdata
  A11 & $ -0.14 \pm 0.17 $ &   1 & $ -0.04 \pm 0.17 $ &   3 & $ -0.22 \pm 0.12 $ &   4 & $ -0.16 \pm 0.10 $ &   5 &    \nodata         &   0\\
  A21 & $ -0.14 \pm 0.17 $ &   1 & $  0.09 \pm 0.18 $ &   3 & $ -0.03 \pm 0.12 $ &   3 & $ -0.14 \pm 0.12 $ &   4 & $ -0.01 \pm 0.16 $ &   1\\
  A25 &    \nodata         &   0 & $  0.09 \pm 0.11 $ &   3 & $  0.04 \pm 0.10 $ &   3 & $ -0.03 \pm 0.12 $ &   4 &    \nodata         &   0\\
  A43 &    \nodata         &   0 & $  0.00 \pm 0.11 $ &   3 & $ -0.07 \pm 0.09 $ &   4 & $ -0.22 \pm 0.11 $ &   7 &    \nodata         &   0\\
  A65 &    \nodata         &   0 & $  0.06 \pm 0.11 $ &   3 & $ -0.12 \pm 0.10 $ &   3 & $ -0.18 \pm 0.12 $ &   4 &    \nodata         &   0\\
   A9 & $ -0.04 \pm 0.17 $ &   1 & $  0.03 \pm 0.17 $ &   3 & $ -0.05 \pm 0.12 $ &   3 & $ -0.11 \pm 0.12 $ &   3 &    \nodata         &   0\\
  A93 &    \nodata         &   0 & $  0.03 \pm 0.10 $ &   4 & $  0.00 \pm 0.09 $ &   4 & $ -0.07 \pm 0.10 $ &   6 &    \nodata         &   0\\
 HB13 & $ -0.21 \pm 0.17 $ &   1 & $  0.27 \pm 0.11 $ &   3 & $  0.18 \pm 0.09 $ &   4 & $  0.04 \pm 0.10 $ &   7 &    \nodata         &   0\\
HB434 &    \nodata         &   0 & $  0.01 \pm 0.14 $ &   3 & $ -0.12 \pm 0.13 $ &   2 & $ -0.31 \pm 0.13 $ &   2 &    \nodata         &   0\\
  HB8 &    \nodata         &   0 & $ -0.02 \pm 0.12 $ &   3 & $  0.02 \pm 0.13 $ &   2 &    \nodata         &   0 &    \nodata         &   0\\
  R21 & $ -0.13 \pm 0.17 $ &   1 & $  0.13 \pm 0.17 $ &   3 & $  0.00 \pm 0.12 $ &   3 & $  0.07 \pm 0.11 $ &   6 & $  0.30 \pm 0.16 $ &   1\\
   R9 & $ -0.15 \pm 0.17 $ &   1 & $  0.05 \pm 0.17 $ &   3 & $ -0.02 \pm 0.12 $ &   3 & $  0.01 \pm 0.12 $ &   3 & $  0.34 \pm 0.16 $ &   1\\
 R100 & $  0.06 \pm 0.17 $ &   1 & $  0.17 \pm 0.12 $ &   3 & $  0.06 \pm 0.12 $ &   2 &    \nodata         &   0 &    \nodata         &   0\\
 R394 &    \nodata         &   0 & $  0.10 \pm 0.11 $ &   3 & $ -0.24 \pm 0.16 $ &   1 &    \nodata         &   0 &    \nodata         &   0\\
 R431 &    \nodata         &   0 & $  0.10 \pm 0.11 $ &   3 & $  0.05 \pm 0.16 $ &   1 &    \nodata         &   0 &    \nodata         &   0\\
  R90 & $ -0.05 \pm 0.17 $ &   1 & $  0.31 \pm 0.17 $ &   3 & $  0.09 \pm 0.17 $ &   1 &    \nodata         &   0 &    \nodata         &   0\\
 I-65 &    \nodata         &   0 & $  0.21 \pm 0.11 $ &   3 & $  0.07 \pm 0.16 $ &   1 &    \nodata         &   0 &    \nodata         &   0
\enddata
\end{deluxetable}

\begin{deluxetable}{lrrrrrrrrrrrr}
\tablecolumns{11}
\tabletypesize{\small}
\tablecaption{Abundance Ratios [NdII/Fe] Through [ThII/Fe] \label{abund6}}
\tablewidth{0pc}
\tablehead{ \colhead{Star ID} & \colhead{[NdII/Fe]} &  \colhead{$N$} & \colhead{[SmII/Fe]} &  \colhead{$N$} & \colhead{[EuII/Fe]} &  \colhead{$N$} & \colhead{[DyII/Fe]} &  \colhead{$N$} & \colhead{[ThII/Fe]} &  \colhead{$N$}  }
\startdata
  A11 & $ -0.08 \pm 0.10 $ &  10 & $  0.14 \pm 0.11 $ &   3 & $  0.37 \pm 0.17 $ &   1 &    \nodata         &   0 & $  0.22 \pm 0.22 $ &   1\\
  A21 & $  0.10 \pm 0.10 $ &   8 & $  0.34 \pm 0.10 $ &   4 & $  0.50 \pm 0.13 $ &   2 &    \nodata         &   0 &    \nodata         &   0\\
  A25 & $  0.10 \pm 0.09 $ &   7 & $  0.31 \pm 0.12 $ &   4 & $  0.47 \pm 0.13 $ &   2 &    \nodata         &   0 &    \nodata         &   0\\
  A43 & $ -0.03 \pm 0.10 $ &  15 & $  0.12 \pm 0.13 $ &   3 & $  0.38 \pm 0.13 $ &   2 & $  0.32 \pm 0.17 $ &   1 &    \nodata         &   0\\
  A65 & $ -0.05 \pm 0.11 $ &   6 & $  0.09 \pm 0.15 $ &   2 & $  0.33 \pm 0.12 $ &   3 & $  0.30 \pm 0.17 $ &   1 &    \nodata         &   0\\
   A9 & $  0.03 \pm 0.11 $ &   8 & $  0.21 \pm 0.13 $ &   2 & $  0.47 \pm 0.13 $ &   2 &    \nodata         &   0 & $  0.29 \pm 0.22 $ &   1\\
  A93 & $  0.08 \pm 0.10 $ &   9 & $  0.17 \pm 0.15 $ &   2 & $  0.38 \pm 0.13 $ &   2 & $  0.43 \pm 0.17 $ &   1 &    \nodata         &   0\\
 HB13 & $  0.18 \pm 0.10 $ &  13 & $  0.31 \pm 0.12 $ &   5 & $  0.52 \pm 0.13 $ &   2 & $  0.60 \pm 0.17 $ &   1 &    \nodata         &   0\\
HB434 & $  0.01 \pm 0.13 $ &   2 &    \nodata         &   0 & $  0.31 \pm 0.17 $ &   1 & $  0.15 \pm 0.17 $ &   1 &    \nodata         &   0\\
  HB8 &    \nodata         &   0 &    \nodata         &   0 & $  0.55 \pm 0.17 $ &   1 &    \nodata         &   0 &    \nodata         &   0\\
  R21 & $  0.12 \pm 0.11 $ &   8 & $  0.33 \pm 0.11 $ &   3 & $  0.52 \pm 0.17 $ &   1 &    \nodata         &   0 & $  0.31 \pm 0.22 $ &   1\\
   R9 & $  0.10 \pm 0.12 $ &   3 & $  0.29 \pm 0.17 $ &   1 & $  0.47 \pm 0.17 $ &   1 &    \nodata         &   0 & $  0.34 \pm 0.22 $ &   1\\
 R100 & $  0.12 \pm 0.14 $ &   2 &    \nodata         &   0 & $  0.51 \pm 0.17 $ &   1 &    \nodata         &   0 &    \nodata         &   0\\
 R394 & $ -0.03 \pm 0.17 $ &   1 &    \nodata         &   0 & $  0.27 \pm 0.17 $ &   1 &    \nodata         &   0 &    \nodata         &   0\\
 R431 & $ -0.02 \pm 0.17 $ &   1 &    \nodata         &   0 & $  0.45 \pm 0.17 $ &   1 &    \nodata         &   0 &    \nodata         &   0\\
  R90 & $  0.18 \pm 0.17 $ &   1 &    \nodata         &   0 & $  0.55 \pm 0.17 $ &   1 &    \nodata         &   0 &    \nodata         &   0\\
 I-65 & $ -0.08 \pm 0.17 $ &   1 &    \nodata         &   0 & $  0.47 \pm 0.17 $ &   1 &    \nodata         &   0 &    \nodata         &   0
\enddata
\end{deluxetable}

\begin{deluxetable}{lrrrrrrrrrrrr}
\tablecolumns{13}
\tabletypesize{\small}
\tablewidth{0pc}
\tablecaption{Average Abundance Values \label{sigma}}
\tablehead{\colhead{Element} & \colhead{[X/Fe]} & \colhead{$\sigma$} & \colhead{N} &
    \colhead{[X/Fe]} & \colhead{$\sigma$} & \colhead{N} & \colhead{[X/Fe]} & \colhead{$\sigma$} & \colhead{N} & \colhead{[X/Fe]} & \colhead{$\sigma$} & \colhead{N} \\
      \colhead{} & \colhead{total} & \colhead{total} & \colhead{total} & \colhead{AGB} & \colhead{AGB} & \colhead{AGB} &
      \colhead{RGB} & \colhead{RGB} & \colhead{RGB} & \colhead{HB} & \colhead{HB} & \colhead{HB}
}
\startdata
       FeI  &  $   -1.51 $  &     0.14  &    17  &  $   -1.53 $  &     0.08  &     8  &  $   -1.40 $  &     0.07  &     7  &  $   -1.80 $  &     0.04  &     2  \\
      FeII  &  $   -1.42 $  &     0.09  &    17  &  $   -1.46 $  &     0.02  &     8  &  $   -1.33 $  &     0.07  &     7  &  $   -1.57 $  &     0.05  &     2  \\
        C   &  $   -0.27 $  &     0.25  &    11  &  $   -0.29 $  &     0.29  &     8  &  $   -0.25 $  &     0.12  &     2  &  $   -0.18 $  &  \nodata  &     1  \\
         N  &  $    0.63 $  &     0.45  &    10  &  $    0.76 $  &     0.40  &     8  &  $    0.14 $  &     0.27  &     2  &    \nodata    &  \nodata  &\nodata \\
         O  &  $    0.31 $  &     0.22  &    15  &  $    0.32 $  &     0.18  &     8  &  $    0.15 $  &     0.13  &     5  &  $    0.65 $  &     0.10  &     2  \\
       Na   &  $    0.19 $  &     0.21  &    17  &  $    0.17 $  &     0.21  &     8  &  $    0.16 $  &     0.24  &     7  &  $    0.34 $  &     0.18  &     2  \\
       Mg   &  $    0.34 $  &     0.07  &    17  &  $    0.31 $  &     0.06  &     8  &  $    0.37 $  &     0.08  &     7  &  $    0.39 $  &     0.06  &     2  \\
       Al   &  $    0.20 $  &     0.25  &    16  &  $    0.17 $  &     0.25  &     8  &  $    0.25 $  &     0.30  &     6  &  $    0.19 $  &     0.13  &     2  \\
       Si   &  $    0.41 $  &     0.08  &    17  &  $    0.42 $  &     0.06  &     8  &  $    0.37 $  &     0.05  &     7  &  $    0.47 $  &     0.21  &     2  \\
       Ca   &  $    0.29 $  &     0.07  &    17  &  $    0.24 $  &     0.06  &     8  &  $    0.32 $  &     0.05  &     7  &  $    0.34 $  &     0.03  &     2  \\
      ScII  &  $    0.15 $  &     0.08  &    17  &  $    0.12 $  &     0.07  &     8  &  $    0.18 $  &     0.08  &     7  &  $    0.15 $  &     0.11  &     2  \\
       TiI  &  $    0.19 $  &     0.06  &    17  &  $    0.15 $  &     0.05  &     8  &  $    0.22 $  &     0.03  &     7  &  $    0.28 $  &     0.04  &     2  \\
      TiII  &  $    0.29 $  &     0.07  &    17  &  $    0.29 $  &     0.07  &     8  &  $    0.30 $  &     0.06  &     7  &  $    0.29 $  &     0.13  &     2  \\
        VI  &  $   -0.15 $  &     0.09  &    16  &  $   -0.22 $  &     0.05  &     8  &  $   -0.07 $  &     0.06  &     7  &  $   -0.17 $  &  \nodata  &     1  \\
       VII  &  $    0.03 $  &     0.11  &     2  &    \nodata    &  \nodata  &\nodata &    \nodata    &  \nodata  &\nodata &  $    0.03 $  &     0.11  &     2  \\
       CrI  &  $   -0.07 $  &     0.08  &    17  &  $   -0.12 $  &     0.08  &     8  &  $   -0.00 $  &     0.03  &     7  &  $   -0.11 $  &     0.09  &     2  \\
      CrII  &  $    0.02 $  &     0.08  &    17  &  $    0.03 $  &     0.05  &     8  &  $    0.02 $  &     0.11  &     7  &  $    0.02 $  &     0.04  &     2  \\
       Mn   &  $   -0.39 $  &     0.06  &    17  &  $   -0.38 $  &     0.04  &     8  &  $   -0.42 $  &     0.06  &     7  &  $   -0.38 $  &     0.11  &     2  \\
       Co   &  $    0.00 $  &     0.09  &    17  &  $   -0.03 $  &     0.09  &     8  &  $    0.04 $  &     0.09  &     7  &  $   -0.02 $  &     0.08  &     2  \\
       Ni   &  $   -0.08 $  &     0.05  &    17  &  $   -0.09 $  &     0.03  &     8  &  $   -0.04 $  &     0.03  &     7  &  $   -0.15 $  &     0.10  &     2  \\
       Zn   &  $    0.16 $  &     0.11  &    10  &  $    0.17 $  &     0.12  &     8  &  $    0.06 $  &  \nodata  &     1  &  $    0.18 $  &  \nodata  &     1  \\
       Cu   &  $   -0.92 $  &     0.06  &    16  &  $   -0.88 $  &     0.04  &     8  &  $   -0.97 $  &     0.05  &     7  &  $   -0.88 $  &  \nodata  &     1  \\
       Sr   &  $   -0.40 $  &     0.05  &     5  &  $   -0.40 $  &     0.05  &     5  &    \nodata    &  \nodata  &\nodata &    \nodata    &  \nodata  &\nodata \\
       YII  &  $   -0.20 $  &     0.10  &    14  &  $   -0.24 $  &     0.07  &     8  &  $   -0.15 $  &     0.06  &     4  &  $   -0.12 $  &     0.24  &     2  \\
       ZrI  &  $    0.10 $  &     0.11  &     7  &  $    0.04 $  &     0.08  &     3  &  $    0.14 $  &     0.11  &     4  &    \nodata    &  \nodata  &\nodata \\
      ZrII  &  $    0.29 $  &     0.12  &     8  &  $    0.32 $  &     0.10  &     7  &    \nodata    &  \nodata  &\nodata &  $    0.08 $  &  \nodata  &     1  \\
       Mo   &  $   -0.10 $  &     0.08  &     8  &  $   -0.13 $  &     0.07  &     4  &  $   -0.07 $  &     0.10  &     4  &    \nodata    &  \nodata  &\nodata \\
      BaII  &  $    0.09 $  &     0.10  &    17  &  $    0.07 $  &     0.09  &     8  &  $    0.15 $  &     0.09  &     7  &  $   -0.00 $  &     0.02  &     2  \\
      LaII  &  $   -0.02 $  &     0.11  &    17  &  $   -0.03 $  &     0.12  &     8  &  $    0.00 $  &     0.11  &     7  &  $   -0.05 $  &     0.10  &     2  \\
      CeII  &  $   -0.10 $  &     0.12  &    11  &  $   -0.11 $  &     0.09  &     8  &  $    0.04 $  &     0.04  &     2  &  $   -0.31 $  &  \nodata  &     1  \\
      PrII  &  $    0.21 $  &     0.19  &     3  &  $   -0.01 $  &  \nodata  &     1  &  $    0.32 $  &     0.03  &     2  &    \nodata    &  \nodata  &\nodata \\
      NdII  &  $    0.05 $  &     0.09  &    16  &  $    0.04 $  &     0.09  &     8  &  $    0.06 $  &     0.10  &     7  &  $    0.01 $  &  \nodata  &     1  \\
      SmII  &  $    0.23 $  &     0.10  &    10  &  $    0.21 $  &     0.10  &     8  &  $    0.31 $  &     0.03  &     2  &    \nodata    &  \nodata  &\nodata \\
      EuII  &  $    0.44 $  &     0.09  &    17  &  $    0.43 $  &     0.07  &     8  &  $    0.46 $  &     0.09  &     7  &  $    0.43 $  &     0.17  &     2  \\
      DyII  &  $    0.36 $  &     0.17  &     5  &  $    0.41 $  &     0.14  &     4  &    \nodata    &  \nodata  &\nodata &  $    0.15 $  &  \nodata  &     1  \\
      ThII  &  $    0.29 $  &     0.05  &     4  &  $    0.25 $  &     0.05  &     2  &  $    0.32 $  &     0.02  &     2  &    \nodata    &  \nodata  &\nodata
\enddata
\end{deluxetable}

\begin{deluxetable}{lrrrr}
\tablecolumns{5}
\tablewidth{0pc}
\tablecaption{Comparisons to Previous Studies for A11 (IV-59)\label{comparison}}
\tablehead{
\colhead{Parameter/} & \colhead{This} & \colhead{This} & \colhead{\citet{ivans01}} &
\colhead{\citet{rc03}} \\
\colhead{Abundance} & \colhead{Study} & \colhead{Study (MARCS)} &\colhead{} & \colhead{} 
}
\startdata
\teff{}           &   4209   &   4209   &   4229  &   4265   \\
\logg{}           &   0.69   &   0.69   &   0.79  &   1.00   \\
$v_t$             &   2.23   &   2.07   &   2.10  &   1.94   \\
~[\ion{Fe}{1}/H]  & $-1.56$  & $-1.60$  & $-1.40$ & $-1.40$  \\
~[\ion{Fe}{2}/H]  & $-1.47$  & $-1.46$  & $-1.25$ & $-1.35$  \\
~[\ion{O}{1}/Fe]  &   0.49   &   0.53   &   0.37  &   0.36   \\
~[\ion{Na}{1}/Fe] &   0.26   &   0.23   &   0.13  &   0.09   \\
~[\ion{Mg}{1}/Fe] &   0.42   &   0.40   & \nodata &   0.29   \\
~[\ion{Si}{1}/Fe] &   0.48   &   0.50   &   0.23  &   0.30   \\
~[\ion{Ca}{1}/Fe] &   0.15   &   0.08   &   0.21  &   0.03   \\
~[\ion{Sc}{2}/Fe] &   0.04   &   0.04   & $-0.19$ &   0.31   \\
~[\ion{Ti}{1}/Fe] &   0.09   & $-0.01$  &   0.08  &   0.01   \\
~[\ion{V}{1}/Fe]  & $-0.23$  & $-0.34$  & $-0.21$ & $-0.29$  \\
~[\ion{Cr}{1}/Fe] & $-0.14$  & $-0.17$  & \nodata & $-0.25$  \\
~[\ion{Mn}{1}/Fe] & $-0.36$  & $-0.41$  & $-0.45$\tablenotemark{*} & $-0.52$  \\
~[\ion{Co}{1}/Fe] &   0.00   & $-0.01$  & \nodata & $-0.13$  \\
~[\ion{Ni}{1}/Fe] & $-0.06$  & $-0.05$  & $-0.14$ & $-0.03$  \\
~[\ion{Cu}{1}/Fe] & $-0.84$  & $-0.90$  & \nodata & $-0.58$  \\
~[\ion{Zn}{1}/Fe] &   0.36   &   0.41   & \nodata &   0.44   \\
~[\ion{Zr}{1}/Fe] &   0.04   & $-0.08$  & \nodata &   0.00   \\
~[\ion{Ba}{2}/Fe] & $-0.04$  & $-0.03$  &   0.10  &   0.19   \\
~[\ion{Eu}{2}/Fe] &   0.37   &   0.36   & \nodata &   0.58   
\enddata
\tablenotetext{*}{From \cite{sobeck06}.}
\end{deluxetable}

\begin{deluxetable}{lcccccc}
\tablecolumns{7}
\tabletypesize{\small}
\tablewidth{0pc}
\tablecaption{[X/Eu] \label{vsEu}}
\tablehead{
\colhead{Element} & \colhead{$\langle$[X/Eu]$\rangle$} & \colhead{$\sigma$} & \colhead{$N$}
& \colhead{[X/Eu]\tablenotemark{*}} &
\colhead{s.s. $r$-process\tablenotemark{*}} & \colhead{s.s $s$-process\tablenotemark{*}}
\\
\colhead{} & \colhead{This Study} & \colhead{} & \colhead{Stars} &
\colhead{s.s. $r$-process} & \colhead{(\%)} & \colhead{(\%)}
}
\startdata
Sr  &  $-0.80$ &   0.14  &   5   & $-0.94$   &  11.0  &  89.0 \\ 
Y   &  $-0.63$ &   0.09  &   14  & $-0.54$   &  28.1  &  71.9 \\
Zr  &  $-0.25$ &   0.14  &   12  & $-0.52$   &  19.1  &  80.9 \\
Mo  &  $-0.59$ &   0.09  &   8   & $-0.48$   &  32.3  &  67.7 \\
Ba  &  $-0.35$ &   0.09  &   17  & $-0.82$   &  14.7  &  85.3 \\
La  &  $-0.46$ &   0.06  &   17  & $-0.60$   &  24.6  &  75.4 \\
Ce  &  $-0.53$ &   0.07  &   11  & $-0.72$   &  18.6  &  81.4 \\
Pr  &  $-0.29$ &   0.20  &   3   & $-0.28$   &  50.8  &  49.2 \\
Nd  &  $-0.39$ &   0.07  &   16  & $-0.36$   &  42.1  &  57.9 \\
Sm  &  $-0.21$ &   0.04  &   10  & $-0.16$   &  66.9  &  33.1 \\
Dy  &  $-0.02$ &   0.10  &   5   & $-0.04$   &  87.9  &  12.1 \\
Th  &  $-0.17$ &   0.04  &   4   & $ 0.02$   & 100.0  &  \nodata 
\enddata
\tablenotetext{*}{From \cite{simmerer04}.}
\end{deluxetable}

\end{document}